\documentclass[10pt,twocolumn]{article}          

\usepackage{authblk}
\providecommand{\keywords}[1]{\textbf{\textit{Keywords---}} #1}
\usepackage[round]{natbib}
\usepackage[top=2cm,bottom=2cm,left=1.5cm,right=1.5cm]{geometry}
\usepackage{tikz}
\usetikzlibrary{patterns}
\usepackage{pifont}
\newcommand{\cmark}{\ding{51}}%
\newcommand{\xmark}{\ding{55}}%
\usepackage{amssymb,amsfonts,amsthm}
\usepackage{mathtools}
\usepackage{algorithm}
\usepackage{algpseudocode}
\usepackage{enumitem}
\usepackage{booktabs}
\usepackage{multirow}
\usepackage{changepage}
\usepackage{hyperref}
\usepackage{titlesec}
 
\titleformat*{\section}{\large\bfseries}
\titleformat*{\subsection}{\bfseries}
\titleformat*{\subsubsection}{\bfseries}
\titleformat*{\paragraph}{\bfseries}
\titleformat*{\subparagraph}{\bfseries}

\theoremstyle{definition}
\newtheorem{definition}{Definition}[section]

\begin{document}

\title{\huge Auxiliary Variables for Bayesian Inference in Multi-Class \\ Queueing Networks \thanks{Work supported by RCUK through the Horizon Digital Economy Research grants (EP/G065802/1, EP/M000877/1) and The Health Foundation through the Insight 2014 project ``Informatics to identify and inform best practice in out of hours secondary care'' (7382).}}
\author[1]{Iker Perez \thanks{Corresponding author, e-mail: iker.perez@nottingham.ac.uk }}
\author[2]{David Hodge}
\author[2]{Theodore Kypraios}
\affil[1]{Horizon Digital Economy Research, University of Nottingham, Nottingham, UK}
\affil[2]{School of Mathematical Sciences, University of Nottingham, Nottingham, UK}
\date{}                     
\setcounter{Maxaffil}{0}
\renewcommand\Affilfont{\itshape\small}

\maketitle

\textit{\small This is a post-peer-review, pre-copy/edit version of an article published in Statistics and Computing. The final authenticated version is available online at: http://dx.doi.org/10.1007/s11222-017-9787-x.
}
\begin{abstract}
Queueing networks describe complex stochastic systems of both theoretical and practical interest. They provide the means to assess alterations, diagnose poor performance and evaluate robustness across sets of interconnected resources. In the present paper, we focus on the underlying continuous-time Markov chains induced by these networks, and we present a flexible method for drawing parameter inference in multi-class Markovian cases with switching and different service disciplines. The approach is directed towards the inferential problem with missing data, where transition paths of individual tasks among the queues are often unknown. The paper introduces a slice sampling technique with mappings to the measurable space of task transitions between the service stations. This can address time and tractability issues in computational procedures, handle prior system knowledge and overcome common restrictions on service rates across existing inferential frameworks. Finally, the proposed algorithm is validated on synthetic data and applied to a real data set, obtained from a service delivery tasking tool implemented in two university hospitals.

\end{abstract}

\section{Introduction} 
Recent literature addressing \textit{queueing networks} (QNs) has aimed to study inferential methods for the estimation of service requirements. These networks offer the means to describe complex stochastic systems through sets of interacting resources, and have found applications in the design of engineering and computing systems \citep{kleinrock1976queueing}, or within call centres \citep{Koole2002}, factories \citep{buzacott1993stochastic} and hospitals \citep{Osorio2009996}. Enabling the understanding of service performance is very important, since it provides quantitative input for the optimal design of interconnected service stations. Yet, drawing inference on parameters is a challenging errand, since in most applications successive network states are never fully observed. Hence, proposed approaches often rely on reduced summaries such as queue lengths, visit counts or response times, and perform inference in different ways, including regression-based estimation procedures, non-linear numerical optimization or maximum likelihood methods.  For a recent review on the matter we refer the reader to \cite{Spinner201551} and references therein. 

In this paper, we focus on the underlying \textit{continuous-time Markov chains} (CTMCs) induced by general-form open QNs, and we develop a flexible framework for drawing Bayesian inference on parameters that govern these models; in the presence of general patterns of missing data currently only discussed in\citep{sutton2011}. Statistical computation is very difficult within this family of models, as it involves working with often countably infinite state spaces where observations provide little indirect information. Here, we target multi-class Markovian cases with possible class switching and different service disciplines, where few or no individual job departure times are observed at specific servers. Hence, knowledge is mostly restricted to task arrival and departures times to, and from, the network. A \textit{task} is a collection of \textit{jobs} undertaken at different service stations, and high loads make it virtually impossible to determine the state of the network at any point in time, including the ordering of jobs across multiple queues. We propose an inferential framework that allows the imposition of prior system knowledge and overcomes common restrictions on service rates across popular service \textit{disciplines} in traditional modelling approaches. A key contribution is that we introduce a slice-sampling approach relying on mappings to the measurable space of task transitions across the service stations; this enables studying systems where the transition paths of tasks among the queues is unknown, and leads to an efficient sampler. The approach draws motivation from techniques aimed to explore countably infinite state spaces within \textit{Dirichlet mixture models} or \textit{infinite-state hidden Markov models} \citep{Walker2007,VanGael2008,Kalli2011}, and sits well within a uniformization oriented MCMC scheme for jump processes as presented in \cite{rao13a}.

Currently, common assumptions in inferential frameworks include the existence of complete data, product-form equilibrium distributions or unique classes with shared \textit{service rates}. However, we often encounter systems where the completion of jobs at individual stations is only occasionally registered. In addition, inference on the basis of balance may in cases be inaccurate; for instance, the existence of equilibrium in service delivery systems with human workers is a strong assumption, since workload is usually externally controlled and arrivals hardly constitute a Poisson process. In addition, there exist concerns regarding the use of steady-state metrics whenever prior knowledge and constraints are imposed on parameters \citep{armero1994assessment}; and the use of product-form solutions within popular \textit{BCMP networks} \citep{baskett1975open} restricts \textit{first come first served} (FCFS) queues to share service distributions over different task classes.

Aiming for flexible inferential methods, Bayesian procedures relying on \textit{Markov Chain Monte Carlo} techniques were first explored in \cite{sutton2011}. There, the authors discussed a latent variable model targeting networks where only subsets of transition times are observed; the method was applicable to open QNs and defined through deterministic transformations between the data and independent \textit{service times} across different disciplines. Later, \cite{Wang2016} proposed the use of a Gibbs sampler relying on product-form distributions and queue length observations, and it advanced the study of closed BCMP networks, offering an approximation method for the normalizing constant within the network's equilibrium distribution. To the best of our knowledge, no further advances exist in the study of exact Monte Carlo inferential frameworks overcoming known restrictions in the study of QNs. Yet, significant progress has been made with sampling techniques and approximate inference methods for continuous-time dynamic systems often modelled as CTMCs  or \textit{continuous-time Bayesian networks} (CTBN) \citep{Nodelman2002,fan2008sampling}. However, simulating system dynamics conditioned on scarce observations remains a complex task; a review on the efficiency of various methods for this purpose (including direct sampling, rejection sampling and uniformization methods) can be found in \cite{hobolth2009}. 

Recently, authors \cite{rao13a} have presented a noteworthy contribution based on the principles of uniformization \citep{Lippman:1975,jensen1953markoff}. Their work explores a class of auxiliary variable MCMC methods allowing for the efficient and exact computation of state evolutions in systems with discrete support (such as Markov jump processes). The framework relies on producing highly dependent time discretizations within subsequent blocked steps in a Gibbs sampler, and is hypothetically applicable to the study of system evolutions within QNs. However, such systems exhibit strong and characteristic temporal dependencies (cf. \cite{sutton2011}), transitions over an infinite set of states, varying specifications of service disciplines and Markovian regimes often subject to switching. Hence, we face major impediments which require elaborate implementations of slice sampling techniques \citep{neal2003}. In this work, we describe a method that controls the computational complexity within simulation procedures; for that matter, we employ families of auxiliary variables across steps in a Gibbs sampler targeting network paths. The result is a method that imposes strong restrictions within the vast space of permissible network transitions at each iteration; however, each subsequent step in the sampler allows for significant timing and routing deviations in limited numbers of tasks routed through the network, ensuring convergence to (i) the distribution of network path evolutions across its full space, given the evidence (ii) the posterior distribution of the arrival and service rates. Finally, we present results on both synthetic and real data, obtained from a service delivery tasking tool implemented in two jointly coordinated university hospitals in the United Kingdom.

The rest of the paper is organised as follows. Section \ref{QNandCTMC} describes CTMCs induced by general form QNs, introduces notions of compatibility with observations, and states the problem addressed in the work. In section \ref{UnifAuxObs} the principle of uniformization and its application to networks is briefly revised, mappings to task transitions and auxiliary variables are introduced and the proposed sampler is described. Section \ref{Examps} introduces results for three example networks of varying complexity with both synthetic and real data. Finally, Section
\ref{Discussion} offers a brief closing discussion.

\section{Queue networks and continuous-time Markov processes} \label{QNandCTMC}

Consider an open Markovian network with $M$ single service stations, a \textit{population set} $\mathcal{C}$ of different task classes and a non-deterministic \textit{network topology} defined by a family of \textit{routing probability matrices} $\mathcal{P} = \{P^c : c\in\mathcal{C}\}$, such that
\begin{itemize}
\item $P^c_{i,j}$ denotes the probability of a class $c\in\mathcal{C}$ task immediately moving to station $j$ after completing a job service in station $i$, for all $1\leq i,j \leq M$.
\item $P^c_{i,0}$ denotes the probability of a class $c\in\mathcal{C}$ task immediately exiting the network after completing a job service in station $i$,  for all $1\leq i\leq M$.
\item $\sum_{j=0}^M P^c_{i,j} = 1$, for all $1\leq i\leq M, c\in\mathcal{C}$.
\end{itemize}
Furthermore, let $\lambda_c>0$ denote external arrival rates for each task class $c\in\mathcal{C}$; and $p^c_{0,i}$ the corresponding probabilities for its first job to enter station $i$, $1\leq i\leq M$. Servers in the network are assumed independent and may differ in their queueing discipline. Service times are non-negative, have constant rates, and vary over servers and classes; we denote them $\mu^c_i$ for all $1\leq i\leq M, c\in\mathcal{C}$. \textit{Switching} is allowed and thus classes are not permanent categorizations; state-dependent service rates are not considered but follow naturally. 

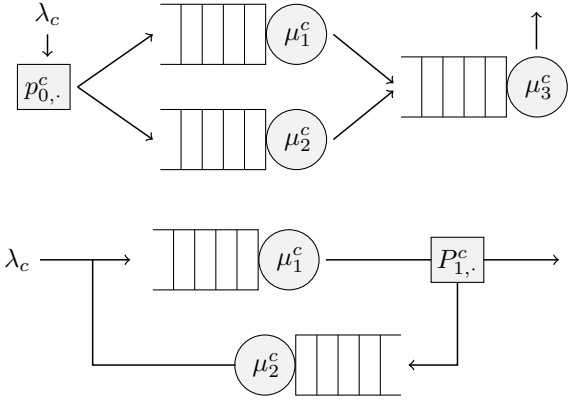
\begin{figure}[h!]
\centering
\begin{tikzpicture}
\draw (0,0) -- ++(1.4cm,0) -- ++(0,-0.8cm) -- ++(-1.4cm,0);
\foreach \i in {1,...,4}
  \draw (1.4cm-\i*8pt,0) -- +(0,-0.8cm);
\filldraw[fill=black!05!white] (1.4cm+0.41cm,-0.4cm) circle [radius=0.4cm]; 

\draw (0,-1.4cm) -- ++(1.4cm,0) -- ++(0,-0.8cm) -- ++(-1.4cm,0);
\foreach \i in {1,...,4}
  \draw (1.4cm-\i*8pt,-1.4cm) -- +(0,-0.8cm);
\filldraw[fill=black!05!white] (1.4cm+0.41cm,-1.4cm-0.4cm) circle [radius=0.4cm]; 

\draw (3.2cm,-0.7cm) -- ++(1.4cm,0) -- ++(0,-0.8cm) -- ++(-1.4cm,0);
\foreach \i in {1,...,4}
  \draw (4.6cm-\i*8pt,-0.7cm) -- +(0,-0.8cm);
\filldraw[fill=black!05!white] (4.6cm+0.41cm,-0.7cm-0.4cm) circle [radius=0.4cm]; 

\filldraw[fill=black!05!white] (-1.9cm,-0.8cm) rectangle (-1.2cm,-1.4cm);

\draw[->,line width=0.20mm] (-1.1cm,-1.1cm) -- (-0.1,-0.4cm);
\draw[->,line width=0.20mm] (-1.1cm,-1.1cm) -- (-0.1,-1.8cm);
\draw[<-,line width=0.20mm] (-1.5cm,-0.7cm) -- +(0cm,0.3cm) node[above] {$\lambda_c$};
\draw[->,line width=0.20mm] (2.3cm,-0.4cm) -- (3.1cm,-1.05cm);
\draw[->,line width=0.20mm] (2.3cm,-1.8cm) -- (3.1cm,-1.15cm);
\draw[->,line width=0.20mm] (5cm,-0.6cm) -- +(0cm,0.5cm);

\node at (1.83,-0.4cm) {$\mu^c_1$};
\node at (1.83,-1.8cm) {$\mu^c_2$};
\node at (5.03,-1.1cm) {$\mu^c_3$};
\node[align=center] at (-1.55cm,-1.15cm) {$p^c_{0,\cdot}$};

\draw (-0.1,-3) -- ++(1.4cm,0) -- ++(0,-0.8cm) -- ++(-1.4cm,0);
\foreach \i in {1,...,4}
  \draw (1.3cm-\i*8pt,-3) -- +(0,-0.8cm);
\filldraw[fill=black!05!white] (1.3cm+0.41cm,-3.4cm) circle [radius=0.4cm]; 

\draw (3.2,-4.4cm) -- ++(-1.4cm,-0cm) -- ++(0cm,-0.8cm) -- ++(1.4cm,0);
\foreach \i in {1,...,4}
  \draw (1.8cm+\i*8pt,-4.4cm) -- +(0,-0.8cm);
\filldraw[fill=black!05!white] (1.8cm-0.41cm,-1.4cm-3.4cm) circle [radius=0.4cm]; 

\filldraw[fill=black!05!white] (3.6cm,-3.1cm) rectangle (4.3cm,-3.7cm);

\draw[-,line width=0.20mm] (2.2cm,-3.4cm) -- ++(1.4,0.0cm);
\draw[->,line width=0.20mm] (4.3cm,-3.4cm) -- ++(1cm,0.0cm);
\draw[->,line width=0.20mm] (3.95cm,-3.7cm) -- ++(0cm,-1.1cm) -- ++(-0.65cm,0cm);
\draw[line width=0.20mm] (1cm,-4.8cm) -- ++(-1.9cm,0cm) -- ++(0cm,1.4cm);
\draw[<-,line width=0.20mm] (-0.4cm,-3.4cm) -- ++(-1.2cm,0cm) node[left] {$\lambda_c$};

\node at (1.73,-3.4cm) {$\mu^c_1$};
\node at (1.41,-4.8cm) {$\mu^c_2$};
\node[align=center] at (3.95cm,-3.45cm) {$P^c_{1,\cdot}$};
\end{tikzpicture}
\vspace{10pt}
\caption{On top, a bottleneck network with $3$ servers; bottom, $2$ networks routed in a loop with a single entry and exit server.} \label{exmplQueues}
\end{figure}

In Figure \ref{exmplQueues} we observe two example networks further examined within Section \ref{Examps} in this paper. There, shaded circles indicate servers with exponential service rates $\mu^c_i$, all accompanied by corresponding job \textit{queueing} areas pictured as empty rectangles. Together, such server and queue pairs each represent a service station $i$, $1\leq i\leq M$. The shaded boxes are probabilistic routing junctions, where task destinations after a job service (or arrival) are determined according to $P^c$ (or $p^c$). Finally, $\lambda_c$ show rates for exponential task arrivals from outside the network.

Under exponential and independence assumptions, there exists an underlying continuous-time Markov process $X=(X_t)_{t\geq 0}$ that describes the system behaviour. Formally, denoting by $\mathcal{S}$ the countably infinite set of possible states in the network, $X$ is a right-continuous stochastic process such that time-indexed variables $X_t$ are defined within a measurable space $(\mathcal{S},\Sigma_\mathcal{S})$, where $\Sigma_\mathcal{S}$ stands for the power set of $\mathcal{S}$. On a basic level, $X$ holds the ordering of jobs in each queue and server, along with their classes and task identifiers; and $\mathcal{S}$ is the multidimensional product of all possible congruent states at every station. The \textit{infinitesimal generator matrix} $Q$ of $X$ is infinite and such that
$$\mathbb{P}(X_{t+\mathrm{d}t}=x'|X_{t}=x) = \mathbb{I}(x=x') + Q_{x,x'}\mathrm{d}t + o(\mathrm{d}t)$$
for all $x,x'\in\mathcal{S}$. Elements in the generator describe rates for transitions within states in the chain, in addition $Q_{x,x'}\geq 0$ for all $x \neq x'$, and $Q_x \coloneqq Q_{x,x} = - \sum_{x'\in\mathcal{S}: x\neq x'} Q_{x,x'}$. Hence, rows in $Q$ sum to $0$, and the full rate for a state departure is given by $|Q_x|$, for all $x\in\mathcal{S}$. Note that transition rates are the product between routing probabilities and exponential rates above; for instance,
\begin{itemize}
\item $\lambda_c p^c_{0,i}$ is the transition rate among states in $\mathcal{S}$ accounting for a class-$c$ arrival to service station $i$,
\item $\mu^c_i P^c_{i,j}$ is the transition rate among states in $\mathcal{S}$ accounting for a job of class $c$ serviced at station $i$ immediately transitioning to station $j$.
\end{itemize}

\subsection{Observations} Let $\Gamma = \{0,\dots,M\}^2 \times \mathbb{N}$ define a \textit{task transition space}. A triplet $\boldsymbol{\gamma}=(i,j,k)\in \Gamma$ denotes a transition for a uniquely identifiable task $k$, with $i$ and $j$ specifying the departure and entry stations respectively. Note that it is possible to augment $\Gamma$ in order to include task classes, yet given unique identifiers this information is redundant. In this work, a transition triplet is never fully observed; instead, we define a set of partial observations $\mathcal{O} = \mathcal{O}_1 \cup \mathcal{O}_ 2 \subset \Sigma_\Gamma$, with
\begin{align*}
\ \mathcal{O}_1 = \{ \sigma \in \Sigma_\Gamma :  \gamma_1=\ &\gamma_1', \ \gamma_3=\gamma_3' \text{ and } \\
 & \gamma_2,\gamma_2'>0, \text{ for all } \boldsymbol{\gamma},\boldsymbol{\gamma}'\in\sigma \}, \\
\ \mathcal{O}_2 = \{ \sigma \in \Sigma_\Gamma :  \gamma_2=\ &\gamma_2'=0, \ \gamma_3=\gamma_3' \text{ and } \\
 & \gamma_1,\gamma_1'>0 , \text{ for all } \boldsymbol{\gamma},\boldsymbol{\gamma}'\in\sigma \}, 
\end{align*}
where $\Sigma_\Gamma$ stands for the power set of $\Gamma$. 
\begin{definition}
An \textit{observation} in $\mathcal{O}$ is a subset of $\Gamma$ that contains all permitted task transitions in the network at some specified time $t>0$, given external information on an arrival, departure or job service.
\end{definition}
\begin{figure}[h]
\centering
\begin{tikzpicture}
\draw[->,line width=0.20mm] (0cm,0cm) -- (7.7cm,0cm);
\draw (0,-0.1cm) -- ++(0cm,0.2cm);
\node at (8cm,0cm) {$\boldsymbol{t}$};
\filldraw[black!60!white,draw=black] (1.2cm,0cm) circle [radius=0.09cm]; 
\filldraw[black!60!white,draw=black] (4.1cm,0cm) circle [radius=0.09cm]; 
\filldraw[black!60!white,draw=black] (6.5cm,0cm) circle [radius=0.09cm]; 
\filldraw[black!05!white,draw=black] (0.75cm,0.82cm) rectangle ++(0.9,0.6);
\filldraw[black!05!white,draw=black] (3.65cm,0.82cm) rectangle ++(0.9,0.6); 
\filldraw[black!05!white,draw=black] (5.5cm,0.82cm) rectangle ++(0.9,0.6);
\filldraw[black!05!white,draw=black] (6.6cm,0.82cm) rectangle ++(0.9,0.6); 
\node[fill=white] at (1.23cm,-0.4cm) {\footnotesize $1_{0\rightarrow 1}$};
\node at (1.2cm,1.1cm) {\small $\boldsymbol{1_{0\rightarrow \cdot}}$};
\node[fill=white] at (4.13cm,-0.4cm) {\footnotesize $1_{1\rightarrow 3}$};
\node at (4.1cm,1.1cm) {\small $\boldsymbol{1_{1\rightarrow \cdot}}$};
\node[fill=white] at (6.53cm,-0.4cm) {\footnotesize $1_{3\rightarrow 0}$};
\node at (5.95cm,1.1cm) {\small $\boldsymbol{1_{3\rightarrow \cdot}}$};
\node at (7.05cm,1.1cm) {\small $\boldsymbol{1_{\cdot\rightarrow 0}}$};
\draw[<-,>=latex,color=black!80!white,line width=0.20mm] (1.2,0.78cm) -- ++(0cm,-0.65cm);
\draw[<-,>=latex,color=black!80!white,line width=0.20mm] (4.1,0.78cm) -- ++(0cm,-0.65cm);
\draw[<-,>=latex,color=black!80!white,line width=0.20mm] (5.95,0.78cm) -- ++(0.5cm,-0.65cm);
\draw[<-,>=latex,color=black!80!white,line width=0.20mm] (7.05,0.78cm) -- ++(-0.5cm,-0.65cm);
\end{tikzpicture}
\caption{Sample observations generated by a single task transitioning a bottleneck network with 3 servers. Observations are represented by rectangles. Dots inform us of transition times. Information below the dots specifies the actual task transitions at each step.} \label{observations}
\end{figure}
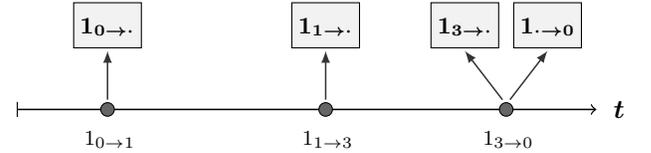

In Figure \ref{observations} we observe a bottleneck network produce four partial observations as it evolves over time. The network corresponds to that in Figure \ref{exmplQueues} (top), and observations include a single task arrival, two job services for the task, and a departure immediately after the final service. There, each task transition $(i,j,k)\in\Gamma$ is marked as $k_{i\rightarrow j}$ at its corresponding time point; note that indexes $i,j$ take the value $0$ in order to specify an external arrival or a departure. The observations take the form of elements of $\mathcal{O}$, i.e.
$$1_{i\rightarrow \cdot} = \{\boldsymbol{\gamma} \in \Gamma : \gamma_1 = i , \gamma_2 >0, \gamma_3 = 1 \}\in\mathcal{O}_1,$$ for $i\in\{1,2,3\}$, and
$$1_{\cdot\rightarrow 0} = \{\boldsymbol{\gamma} \in \Gamma : \gamma_1 > 0 , \gamma_2 =0, \gamma_3 = 1 \}\in\mathcal{O}_2.$$
In this toy example, it is possible to deduce the original path $X$ in the network when considering the available observations along with the topology in Figure \ref{exmplQueues}; including task orderings across all queues and servers at every point in time. However, in real world applications job service observations are often missing or do not exist at all. In this work, only arrivals and departures are assumed to always be available.

\subsection{Compatibility}

Let  $\mathcal{T}: \mathcal{S}^2 \rightarrow \Gamma \cup \varnothing$ define a measurable function, equipped with the corresponding products of discrete algebras, which maps a pair of states $x,x'\in\mathcal{S}$ to its task transition triplet in $\Gamma$. For instance, $$\mathcal{T}(x,x')=(2,3,12)$$ should $x'$ be reachable from $x$ by servicing a job for task $12$ in server $2$ and immediately routing it to queue $3$. Note that for this to be possible, a job for task $12$ must be in server $2$ within $x$, and the remaining tasks in the system must be distributed and ordered across stations so that there will exist full agreement with $x'$. If a state $x'$ is not directly reachable from $x$, then $\mathcal{T}(x,x') = \varnothing$. We note that the pre-image of a triplet in $\mathcal{T}$ is given by a countably infinite set of pairs of network states in $\mathcal{S}$, unless bounds on the task population are imposed. 

\begin{definition} \label{comp} Fix some terminal time $T>0$ and let $\{O_{t_r}\in\mathcal{O} : r=1,\dots,R\}$ be a sequence of observations at times $0\leq t_1<\dots<t_R\leq T$. Also, let $Y_{t_r} = \mathcal{T}^{-1}(O_{t_r}) \in \Sigma_\mathcal{S}^2$ for all $r=1,\dots,R$. Then, we say that a process $X$ is \textit{compatible} with an observation $O_{t_r}$, and we write $X\perp O_{t_r}$ if
$$ \lim_{s\nearrow t_r}X_s = y \quad \text{and} \quad X_{t_r} = y', $$
for some pair of network configurations $(y,y') \in Y_{t_r}$. Furthermore, we say that a process $X$ is \textit{fully compatible} with the observations if $X\perp O_{t_r}$ for all $r=1,\dots,R$.
\end{definition}

\begin{figure}[h]
\centering
\begin{tikzpicture}
\draw[->,line width=0.20mm] (0cm,0cm) -- (7.7cm,0cm);
\draw (0,-0.1cm) -- ++(0cm,0.2cm);
\draw[->,line width=0.20mm] (0cm,-1.5cm) -- (7.7cm,-1.5cm);
\draw (0,-1.6cm) -- ++(0cm,0.2cm);
\draw[->,line width=0.20mm] (0cm,-3cm) -- (7.7cm,-3cm);
\draw (0,-3.1cm) -- ++(0cm,0.2cm);
\draw[-,dashed,color=black!80!white,line width=0.20mm] (1.2,0.68cm) -- ++(0cm,-4.2cm);
\draw[-,dashed,color=black!80!white,line width=0.20mm] (6.5,0.68cm) -- ++(0cm,-4.2cm);
\node at (8cm,0cm) {$\boldsymbol{t}$};
\node at (8cm,-1.5cm) {$\boldsymbol{t}$};
\node at (8cm,-3cm) {$\boldsymbol{t}$};
\filldraw[black!60!white,draw=black] (1.2cm,0cm) circle [radius=0.09cm]; 
\filldraw[black!60!white,draw=black] (4.1cm,0cm) circle [radius=0.09cm]; 
\filldraw[black!60!white,draw=black] (6.5cm,0cm) circle [radius=0.09cm]; 
\filldraw[black!60!white,draw=black] (1.2cm,-1.5cm) circle [radius=0.09cm]; 
\filldraw[black!60!white,draw=black] (2.8cm,-1.5cm) circle [radius=0.09cm]; 
\filldraw[black!60!white,draw=black] (6.5cm,-1.5cm) circle [radius=0.09cm]; 
\filldraw[black!60!white,draw=black] (1.2cm,-3cm) circle [radius=0.09cm]; 
\filldraw[black!60!white,draw=black] (4.9cm,-3cm) circle [radius=0.09cm]; 
\filldraw[black!60!white,draw=black] (6.5cm,-3cm) circle [radius=0.09cm]; 
\filldraw[black!05!white,draw=black] (0.75cm,0.8cm) rectangle ++(0.9,0.6);
\filldraw[black!05!white,draw=black] (6.05cm,0.8cm) rectangle ++(0.9,0.6); 
\node[fill=white] at (1.23cm,-0.4cm) {\footnotesize $1_{0\rightarrow 1}$};
\node at (1.2cm,1.08cm) {\small $\boldsymbol{1_{0\rightarrow \cdot}}$};
\node[fill=white] at (4.13cm,-0.4cm) {\footnotesize $1_{1\rightarrow 3}$};
\node[fill=white] at (6.53cm,-0.4cm) {\footnotesize $1_{3\rightarrow 0}$};
\node at (6.5cm,1.08cm) {\small $\boldsymbol{1_{\cdot\rightarrow 0}}$};

\node[fill=white] at (1.23cm,-1.9cm) {\footnotesize $1_{0\rightarrow 2}$};
\node[fill=white] at (2.83cm,-1.9cm) {\footnotesize $1_{2\rightarrow 3}$};
\node[fill=white] at (6.53cm,-1.9cm) {\footnotesize $1_{3\rightarrow 0}$};

\node[fill=white] at (1.23cm,-3.4cm) {\footnotesize $1_{0\rightarrow 2}$};
\node[fill=white] at (4.93cm,-3.4cm) {\footnotesize $1_{2\rightarrow 3}$};
\node[fill=white] at (6.53cm,-3.4cm) {\footnotesize $1_{3\rightarrow 0}$};
\end{tikzpicture}
\caption{Example network paths, all compatible with arrival and departure observations for a single task entering and leaving a bottleneck network with three servers.} \label{compatibilityPaths}
\end{figure}
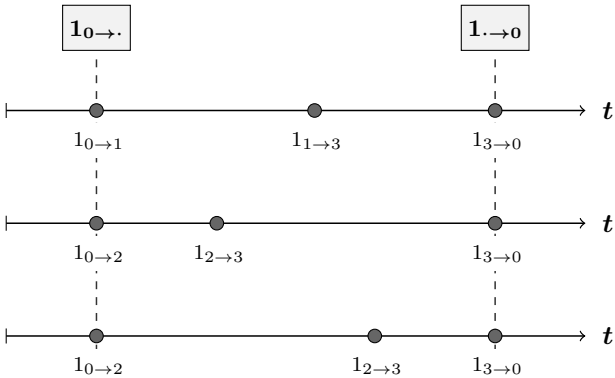
In Figure \ref{compatibilityPaths} we observe task transitions for sample paths $X$ which are compatible with the arrival and departure information as shown in Figure \ref{observations}. There, notice that the first sequence corresponds to the original path forming the observations. This time, no job services have been retained and there exist infinitely many paths $X$ that could have produced the same output, with varying transition times and task orderings across the different stations. In large networks with multiple tasks and all simultaneously transitioning the system, it is hard to picture the infinite amount of fully compatible paths $X$, unless large proportions of job services are retrieved.

\subsection{Latent network and problem statement} Denote by $x_0\in\mathcal{S}$ the initial state in $X$. In this paper, this is assumed to be an \textit{empty} state, where no jobs populate the network. It is however possible to define an initial distribution $\pi$ over states, s.t. $\pi(x)\coloneqq \mathbb{P}(X_0=x)$ for all $x\in\mathcal{S}$. Now, assume we retrieve $K\in\mathbb{N}$ observation sequences $\tilde{\boldsymbol{O}}=\{\boldsymbol{O}_k\}_{k=1,\dots,K}$ collected during different realizations $\boldsymbol{X}=\{X^k\}_{k=1,\dots,K}$ in the network; with 
$$\boldsymbol{O}_k = \{O_{t_r}\in\mathcal{O} : r=1,\dots,R_k\}$$
at times $0\leq t_1<\dots<t_{R_k}\leq T_k$, for $k=1,\dots,K$.

The likelihood function is proportional to the product of network path densities fully compatible with $\tilde{\boldsymbol{O}}$, and is thus intractable. A Gibbs sampling approach centred around latent network evolutions is appropriate, iterating between paths and parameters. For that, note that every $X$ is a piecewise-constant process and may be fully characterized by a set of transition times $\boldsymbol{t}=\{t_1,\dots,t_{n}\}$ along with network states $\boldsymbol{x}=\{x_1,\dots,x_{n}\}$, so that $X\equiv (\boldsymbol{t},\boldsymbol{x})$ with $X_0 = x_0$. To ease notation, denote $\boldsymbol{\theta} \equiv (\mathcal{P},\boldsymbol{p},\boldsymbol{\lambda},\boldsymbol{\mu})$, where $\boldsymbol{p}$ is the vector of arrival routing probabilities. Now, let $\delta_{\boldsymbol{x}}$ be the number of transitions in $\boldsymbol{x}$ excluding task arrivals and departures. For each $k=1\dots,K$, the density of $(\boldsymbol{t},\boldsymbol{x})$ given $\boldsymbol{O}_k$ is (up to proportionality) such that
\begin{align}
f_{X}((\boldsymbol{t},\boldsymbol{x})|&\boldsymbol{\theta},\boldsymbol{O}_k,x_0)  \nonumber \\
& \propto f_{\boldsymbol{O}}(\boldsymbol{O}_k|(\boldsymbol{t},\boldsymbol{x}),x_0) f_{X}((\boldsymbol{t},\boldsymbol{x})|\boldsymbol{\theta},x_0) \nonumber \\
& \propto (1-q)^{\delta_{\boldsymbol{x}}-\delta_{\boldsymbol{o}}} \times \mathbb{I}((\boldsymbol{t},\boldsymbol{x}) \perp O : O\in\boldsymbol{O}_k) \nonumber \\
& \times e^{Q_{x_n}(T_k-t_n)} \prod_{i=1}^n Q_{x_{i-1},x_{i}} e^{Q_{x_{i-1}}(t_i-t_{i-1})}, \label{densPath}
\end{align}
where $q$ denotes the probability that a job service in $X$ is observed, and $\delta_{\boldsymbol{o}}\leq \delta_{\boldsymbol{x}}$ is the corresponding amount of service observations in $\boldsymbol{O}_k$. This density is supported in a suitably defined space of finite network evolutions and the term on top is proportional to Bernoulli trials penalizing network paths with unobserved job services. The term below follows from the properties of the minimum of exponentially distributed random variables. 

Hence, drawing parameter inference entails the complex task of simulating network configurations from \eqref{densPath}, over increasingly large state spaces and with strong conditional dependencies.  In the following, we revise the notion of uniformization and sampling methods for jump processes introduced in \cite{rao13a}, and we present an auxiliary observation-variable sampler fit for inference in QN models.

\section{Uniformization and auxiliary observations} \label{UnifAuxObs}

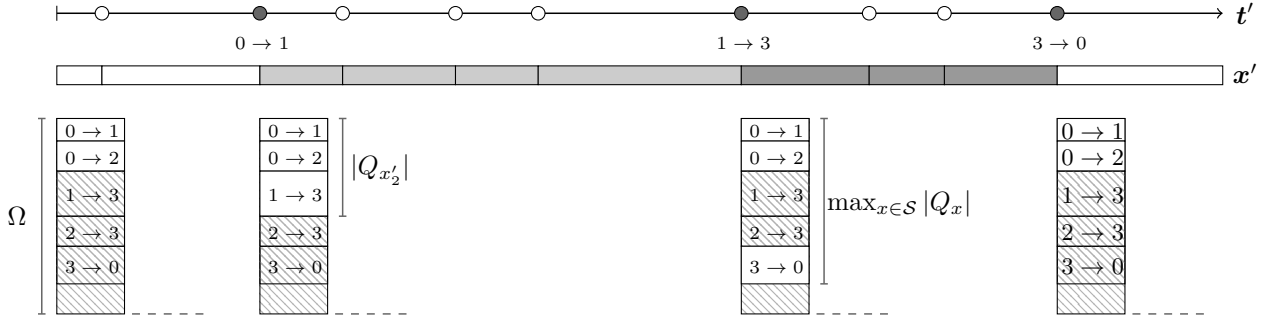
\begin{figure*}[t]
\centering
\begin{tikzpicture}

\draw[->,line width=0.20mm] (0cm,0cm) -- (15.5cm,0cm);
\draw (0,-0.1cm) -- ++(0cm,0.2cm);

\node at (15.8cm,0cm) {$\boldsymbol{t}'$};

\filldraw[black!00!white,draw=black] (0.6cm,0cm) circle [radius=0.09cm]; 
\filldraw[black!60!white,draw=black] (2.7cm,0cm) circle [radius=0.09cm]; 
\filldraw[black!00!white,draw=black] (3.8cm,0cm) circle [radius=0.09cm]; 
\filldraw[black!00!white,draw=black] (5.3cm,0cm) circle [radius=0.09cm]; 
\filldraw[black!00!white,draw=black] (6.4cm,0cm) circle [radius=0.09cm]; 
\filldraw[black!60!white,draw=black] (9.1cm,0cm) circle [radius=0.09cm]; 
\filldraw[black!00!white,draw=black] (10.8cm,0cm) circle [radius=0.09cm]; 
\filldraw[black!00!white,draw=black] (11.8cm,0cm) circle [radius=0.09cm]; 
\filldraw[black!60!white,draw=black] (13.3cm,0cm) circle [radius=0.09cm]; 

\node[fill=white] at (2.73cm,-0.4cm) {\scriptsize $0\rightarrow 1$};
\node[fill=white] at (9.13cm,-0.4cm) {\scriptsize $1\rightarrow 3$};
\node[fill=white] at (13.33cm,-0.4cm) {\scriptsize $3\rightarrow 0$};

\filldraw[draw=black,fill=black!00!white] (0.0,-0.7) rectangle (0.6,-0.95);
\filldraw[draw=black,fill=black!00!white] (0.6,-0.7) rectangle (2.7,-0.95);
\filldraw[draw=black,fill=black!20!white] (2.7,-0.7) rectangle (3.8,-0.95);
\filldraw[draw=black,fill=black!20!white] (3.8,-0.7) rectangle (5.3,-0.95);
\filldraw[draw=black,fill=black!20!white] (5.3,-0.7) rectangle (6.4,-0.95);
\filldraw[draw=black,fill=black!20!white] (6.4,-0.7) rectangle (9.1,-0.95);
\filldraw[draw=black,fill=black!40!white] (9.1,-0.7) rectangle (10.8,-0.95);
\filldraw[draw=black,fill=black!40!white] (10.8,-0.7) rectangle (11.8,-0.95);
\filldraw[draw=black,fill=black!40!white] (11.8,-0.7) rectangle (13.3,-0.95);
\filldraw[draw=black,fill=black!00!white] (13.3,-0.7) rectangle (15.5,-0.95);

\node at (15.8cm,-0.8cm) {$\boldsymbol{x}'$};

\draw[draw=black,pattern=north west lines,pattern color=black!30!white] (0.0,-1.4) rectangle (0.9,-4cm);
\filldraw[draw=black,fill=black!00!white] (0.0,-1.4) rectangle (0.9,-1.7);
\node at (0.47cm,-1.57cm) {\scriptsize $0\rightarrow 1$};
\filldraw[draw=black,fill=black!00!white] (0.0,-1.7) rectangle (0.9,-2.1);
\node at (0.47cm,-1.92cm) {\scriptsize $0\rightarrow 2$};
\draw[draw=black,pattern=north west lines,pattern color=black!30!white] (0.0,-2.1) rectangle (0.9,-2.7);
\node at (0.47cm,-2.42cm) {\scriptsize $1\rightarrow 3$};
\draw[draw=black,pattern=north west lines,pattern color=black!30!white] (0.0,-2.7) rectangle (0.9,-3.1);
\node at (0.47cm,-2.92cm) {\scriptsize $2\rightarrow 3$};
\draw[draw=black,pattern=north west lines,pattern color=black!30!white] (0.0,-3.1) rectangle (0.9,-3.6);
\node at (0.47cm,-3.37cm) {\scriptsize $3\rightarrow 0$};

\draw[draw=black,pattern=north west lines,pattern color=black!30!white] (2.7,-1.4) rectangle (3.6,-4cm);
\filldraw[draw=black,fill=black!00!white] (2.7,-1.4) rectangle (3.6,-1.7);
\node at (2.7 + 0.47,-1.57cm) {\scriptsize $0\rightarrow 1$};
\filldraw[draw=black,fill=black!00!white] (2.7,-1.7) rectangle (3.6,-2.1);
\node at (2.7 + 0.47,-1.92cm) {\scriptsize $0\rightarrow 2$};
\filldraw[draw=black,fill=black!00!white] (2.7,-2.1) rectangle (3.6,-2.7);
\node at (2.7 + 0.47,-2.42cm) {\scriptsize $1\rightarrow 3$};
\draw[draw=black,pattern=north west lines,pattern color=black!30!white] (2.7,-2.7) rectangle (3.6,-3.1);
\node at (2.7 + 0.47,-2.92cm) {\scriptsize $2\rightarrow 3$};
\draw[draw=black,pattern=north west lines,pattern color=black!30!white] (2.7,-3.1) rectangle (3.6,-3.6);
\node at (2.7 + 0.47,-3.37cm) {\scriptsize $3\rightarrow 0$};

\draw[draw=black,pattern=north west lines,pattern color=black!30!white] (9.1,-1.4) rectangle (10,-4cm);
\filldraw[draw=black,fill=black!00!white] (9.1,-1.4) rectangle (10,-1.7);
\node at (9.1 + 0.47,-1.57cm) {\scriptsize $0\rightarrow 1$};
\filldraw[draw=black,fill=black!00!white] (9.1,-1.7) rectangle (10,-2.1);
\node at (9.1 + 0.47,-1.92cm) {\scriptsize $0\rightarrow 2$};
\draw[draw=black,pattern=north west lines,pattern color=black!30!white] (9.1,-2.1) rectangle (10,-2.7);
\node at (9.1 + 0.47,-2.42cm) {\scriptsize $1\rightarrow 3$};
\draw[draw=black,pattern=north west lines,pattern color=black!30!white] (9.1,-2.7) rectangle (10,-3.1);
\node at (9.1 + 0.47,-2.92cm) {\scriptsize $2\rightarrow 3$};
\filldraw[draw=black,fill=black!00!white] (9.1,-3.1) rectangle (10,-3.6);
\node at (9.1 + 0.47,-3.37cm) {\scriptsize $3\rightarrow 0$};

\draw[draw=black,pattern=north west lines,pattern color=black!30!white] (13.3,-1.4) rectangle (14.2,-4cm);
\filldraw[draw=black,fill=black!00!white] (13.3,-1.4) rectangle (14.2,-1.7);
\node at (13.3 + 0.47,-1.57cm) {\small $0\rightarrow 1$};
\filldraw[draw=black,fill=black!00!white] (13.3,-1.7) rectangle (14.2,-2.1);
\node at (13.3 + 0.47,-1.92cm) {\small $0\rightarrow 2$};
\draw[draw=black,pattern=north west lines,pattern color=black!30!white] (13.3,-2.1) rectangle (14.2,-2.7);
\node at (13.3 + 0.47,-2.42cm) {\small $1\rightarrow 3$};
\draw[draw=black,pattern=north west lines,pattern color=black!30!white] (13.3,-2.7) rectangle (14.2,-3.1);
\node at (13.3 + 0.47,-2.92cm) {\small $2\rightarrow 3$};
\draw[draw=black,pattern=north west lines,pattern color=black!30!white] (13.3,-3.1) rectangle (14.2,-3.6);
\node at (13.3 + 0.47,-3.37cm) {\small $3\rightarrow 0$};

\draw[-,color=black!60!white,line width=0.20mm] (-0.2cm,-1.4 cm) -- (-0.2cm,-4cm);
\draw[-,color=black!60!white,line width=0.20mm] (-0.25cm,-1.4 cm) -- (-0.15cm,-1.4cm);
\draw[-,color=black!60!white,line width=0.20mm] (-0.25cm,-4 cm) -- (-0.15cm,-4cm);
\node at (-0.53,-2.7) {$\Omega$};

\draw[-,color=black!60!white,line width=0.20mm] (10.2cm,-1.4 cm) -- (10.2cm,-3.6cm);
\draw[-,color=black!60!white,line width=0.20mm] (10.25cm,-1.4 cm) -- (10.15cm,-1.4cm);
\draw[-,color=black!60!white,line width=0.20mm] (10.25cm,-3.6 cm) -- (10.15cm,-3.6cm);
\node at (11.2,-2.53) {$\max_{x\in\mathcal{S}}|Q_x|$};

\draw[-,color=black!60!white,line width=0.20mm] (3.8cm,-1.4 cm) -- (3.8cm,-2.7cm);
\draw[-,color=black!60!white,line width=0.20mm] (3.75cm,-1.4 cm) -- (3.85cm,-1.4cm);
\draw[-,color=black!60!white,line width=0.20mm] (3.75cm,-2.7 cm) -- (3.85cm,-2.7cm);
\node at (4.3,-2.08) {$|Q_{x_2'}|$};

\draw[-,dashed,color=black!60!white,line width=0.20mm] (1cm,-4 cm) -- (2cm,-4cm);
\draw[-,dashed,color=black!60!white,line width=0.20mm] (3.7cm,-4 cm) -- (4.7cm,-4cm);
\draw[-,dashed,color=black!60!white,line width=0.20mm] (10.1cm,-4 cm) -- (11.1cm,-4cm);
\draw[-,dashed,color=black!60!white,line width=0.20mm] (14.3cm,-4 cm) -- (15.3cm,-4cm);
\end{tikzpicture}
\vspace{10pt}
\caption{Graphical representation of times $\boldsymbol{t}'$, states $\boldsymbol{x}'$ and transition probabilities for a uniformization-based simulation in a single-class bottleneck network. We observe a single task routed from entry to departure, with virtual transitions represented by empty dots. Vertical rectangles are proportionally split according the likelihood of the various possible services and arrivals.} \label{unifQueue}
\end{figure*}

A generative process for sampling $X$ requires alternating between exponentially distributed times and transitions in proportion to rates. Instead, a \textit{uniformization}-based \citep{Lippman:1975,jensen1953markoff} sampling scheme employs a \textit{dominating rate} $\Omega$ and introduces the notion of \textit{virtual transitions}, so that all times are sampled in a blocked step. In Algorithm \ref{unifProcX} we observe a uniformization procedure to produce network paths.
\begin{algorithm}[h!]
  \normalsize
\caption{Uniformization procedure for process $X$}
\label{unifProcX}
\begin{algorithmic}[1] \vspace{2pt}
\State Fix a dominating rate $\Omega \geq \max_{x\in\mathcal{S}}|Q_x|$. 
\State Sample transition times $0\leq t_1< \dots <t_m\leq T$ from a homogeneous Poisson process with rate $\Omega$. 
\State Set initial state $X_0=x_0$. 
\For {$i\in\{1,\dots,m\}$}
	\State Sample $x_i$ from $\{x_{i-1}\}\cup \mathcal{X}_{x_{i-1}}$, with $$\mathcal{X}_{x_{i-1}}=\{x\in\mathcal{S}\backslash\{x_{i-1}\} : \mathcal{T}(x_{i-1},x)\neq\varnothing\},$$  \hskip\algorithmicindent and probabilities $\pi_{x_{i-1}}$ given by $$\pi_{x_{i-1}} = \{1+Q_{x_{i-1}}/\Omega\} \cup \{Q_{x_{i-1},x}/\Omega : x\in\mathcal{X}_{x_{i-1}}\}.$$
\EndFor
\State Return $(\boldsymbol{t},\boldsymbol{x})$.
\end{algorithmic}
\end{algorithm}

A proof of probabilistic equivalence between a generative and uniformized sampling scheme involves comparing the marginal distribution across states at any time $t\geq 0$, and can be found in \cite{hobolth2009}. A uniformization procedure yields an augmented set of times $\boldsymbol{t}'=\{t_1',\dots,t_{m}'\}$ and states $\boldsymbol{x}'=\{x_1',\dots,x_{m}'\}$ that accounts for both real and virtual transitions in $X$. Whenever $x_i=x_{i-1}$ we refer to transition $i$ as \textit{virtual} and note that the number of such transitions is dependent on the choice of $\Omega$. Finally, the density function in \eqref{densPath} may be rewritten to include virtual jumps, so that
\begin{align*}
f_{X}((\boldsymbol{t}',\boldsymbol{x}')|\Omega&,\boldsymbol{\theta},\boldsymbol{O}_k,x_0) \\
&\propto (1-q)^{\delta_{\boldsymbol{x}'}-\delta_{\boldsymbol{o}}} \times \mathbb{I}((\boldsymbol{t}',\boldsymbol{x}') \perp O : O\in\boldsymbol{O}_k) \nonumber \\
& \times \prod_{i=1}^m Q_{x_{i-1}',x_{i}'}^{\mathbb{I}(x_i'\neq x_{i-1}')} (\Omega+Q_{x_{i-1}'})^{\mathbb{I}(x_i'= x_{i-1}')},
\end{align*} 
where terms not proportional to $(\boldsymbol{t}',\boldsymbol{x}')$ are omitted.

In practice, simulating $X$ only requires considering a limited number of candidate states in each transition; in close relation to the number of service stations. In Figure \ref{unifQueue} we observe a graphical representation of times, states and transition probabilities for a uniformization-based procedure in the single-class bottleneck network in Figure \ref{exmplQueues} (top). There, we observe only one task from entry to departure, and we notice $\boldsymbol{x}'$ is unaltered after virtual transitions. Vertical rectangles are divided in proportion to rates for services and arrivals, and infeasible services are hashed in grey (the additional hashed area in the bottom accounts for a strictly positive dominating rate $\Omega$). This determines the probabilities leading to new states at subsequent times, with virtual jumps associated to the sum of all hashed regions. Finally, removing virtual transitions within $(\boldsymbol{t}',\boldsymbol{x}')$ yields the desired realization $(\boldsymbol{t},\boldsymbol{x})$ in $X$.

\subsection{An auxiliary observation-variable sampler} A uniformization oriented approach can enable the construction of a Gibbs sampler targeting the conditional distribution $f_{X}((\boldsymbol{t},\boldsymbol{x})|\boldsymbol{\theta},\boldsymbol{O}_k,x_0)$. For such purpose, authors \cite{rao13a} show it is possible to recycle groups of real transition times within each iteration. The method applies well to many families of Markov jump processes, but it is insufficient in order to tackle complex systems such as QNs due to a quadratic cost on the number of states when producing $\boldsymbol{x}$. This is a known problem in discrete-time systems with large state spaces (such as dynamic Bayesian networks or infinite-state hidden Markov models), and proposed solutions include approximate inference methods \citep{Boyen1998,Ng2002} or the use of slice sampling techniques for exact inference \citep{VanGael2008}. 

However, QNs contain strong serial dependencies, and transitions over an infinite set of states are triggered by a very reduced number of rates; hence, this can render techniques aimed at Dirichlet mixture models \citep{Walker2007,Kalli2011} or hidden Markov models unusable. A viable approach would ideally consider limited divergences in network paths $X$ over subsequent steps in a sampler; yet allowing for considerable deviations in the routing of a reduced set of tasks. Here, we describe a sampling scheme that achieves this goal by employing random auxiliary mappings to the space of task transitions $\Gamma$. Intuitively:
\begin{itemize}
\item In each iteration we will first produce additional auxiliary evidence, resulting from task transitions within the current trajectory of $X$. 
\item This evidence will be used next in order to significantly restrict the explorable range of network paths in the the following sampler iteration.
\end{itemize}
This approach poses a computationally tractable technique suited for the analysis of system transitions in QNs, and will construct a Markov chain of posterior trajectories over the entire range of paths in full agreement with the original observed evidence, where reasonably distant samples in the chain are statistically unrelated.

\subsubsection{Preliminaries}

Set $\Omega > \max_{x\in\mathcal{S}}|Q_x|$ and let $\boldsymbol{t}'$ and $\boldsymbol{x}'$ define some auxiliary frames of transition times and states in $X$, including both real and virtual values.  Arrival, departure and job service observations must come at transition times in $\boldsymbol{t}'$; hence, we may define an augmented set of observations $\boldsymbol{O}_k' = \{O_{t_i'}\in\mathcal{O}\cup\mathcal{O}_3 : i=1,\dots,m\}$ at times $t_1',\dots,t_m'$, with
$$\mathcal{O}_3 = \{ \{(i,j,k)\in\Gamma : i,j\neq 0 \}\cup\{\varnothing\} \}$$
and such that $\boldsymbol{O}_k\subseteq\boldsymbol{O}_k'$. This accounts for missing observations; note that since arrivals and departures are always observed, a missing observation offers evidence for either an inner transition or virtual jump in the network. For simplicity, we assume that no state is reachable from itself in a transition, so that $\mathcal{T}(x,x)=\varnothing$; however, the framework naturally extends to networks where self-transitions are a possibility. Now, denote by $\boldsymbol{u}$ an auxiliary family of subsets of $\Gamma\cup\varnothing$, such that
\begin{align}
\  \mathbb{P}(u_i=u|x_{i-1}',x_{i}') = \begin{cases}
       p, & \text{if } u= \{\mathcal{T}(x_{i-1}',x_{i}')\}, \\
        1-p, & \text{if }u=\Gamma\cup\varnothing,
        \end{cases} \label{auxes}
\end{align}
with some fixed $p\in[0,1]$, for all $u_i\in\boldsymbol{u}$, $i=1,\dots,m$. Hence, auxiliary variables $u\in\boldsymbol{u}$ will refer to either the entire space of task transitions or sets with a single element in $\Gamma$; we note that these single element sets will be further contained within a larger observation-set $O\in\boldsymbol{O}_k'$. 

Recall that in queueing networks a task transition may follow from an infinite number of network configurations; that is, there may exist an infinite amount of task orderings across the stations so that a specific job is serviced in one given server and routed to another. However, any network state can only transition to a finite space, by relocating one task in a new queue after a service or an arrival. Thus $$\{(x,x')\}\subset\mathcal{T}^{-1}\mathcal{T}(x,x')\subset \mathcal{S}^2,$$ strictly, for all $x,x'\in\mathcal{S}$. Moreover, any $u\in\boldsymbol{u}$ such that $u\neq \Gamma\cup\varnothing$ can only be produced by a limited set of uniformized paths in $X$, and compatibility definitions in Definition \ref{comp} extend naturally to these auxiliary-observation variables. Restrictions are of two types:
\begin{itemize}
\item \textit{Transition triplets} impose a transition for an identifiable task. The transition probability is identical over all pairs of compatible states $(x,x')\in\mathcal{S}$.
\item \textit{Null sets} impose virtual jumps. The transition probability (lack thereof) depends both on the network configuration and dominating rate $\Omega$.
\end{itemize}

\subsubsection{Sampler} 
\begin{algorithm*}[t]
\normalsize
\caption{Forward filtering backward sampling with dynamic arrays}
\label{backForwAlgo}
\begin{algorithmic}[1] \vspace{3pt}
\State Set initial state $x_0$ and let $\alpha_0(x)=\mathbb{I}(x= x_0)$, $x\in\mathcal{S}$.  
\For {$i\in\{1,\dots,m\}$} \hfill \textit{Forward Filtering}
	\For {$x\in\mathcal{S}$ s.t. $\alpha_{i-1}(x)>0$}
		\For {$x'\in\mathcal{S}$ s.t. $|Q_{x,x'}|>0,(x,x')\in\mathcal{T}^{-1}(O_{t_i'}) \cap \mathcal{T}^{-1}(u_i)$}
			\State Update: $$\alpha_{i}(x') \gets  \alpha_{i}(x') + (1-q)^{\mathbb{I}(\mathcal{T}(x,x')\neq\varnothing)} \Big (\mathbb{I}(x=x')+ \frac{Q_{x,x'}}{\Omega}\Big) \alpha_{i-1}(x)$$
		\EndFor
	\EndFor
\EndFor
\State Sample $x_m$ from $x\in\mathcal{S}$ with probability in proportion to $\alpha_m(x)$.
\For {$i\in\{m-1,\dots,1\}$}   \hfill \textit{Backward Sampling}
	\For {$x\in\mathcal{S}$ s.t. $|Q_{x,x_{i+1}}|>0,\alpha_i(x)>0,(x,x_{i+1})\in \mathcal{T}^{-1}(O_{t_{i+1}'}) \cap \mathcal{T}^{-1}(u_{i+1})$} 
		\State Update: $$\beta(x) \gets \alpha_i(x) (1-q)^{\mathbb{I}(\mathcal{T}(x,x_{i+1})\neq\varnothing)} \Big(\mathbb{I}(x=x_{i+1})+ \frac{Q_{x,x_{i+1}}}{\Omega}\Big)$$
	\EndFor
	\State Sample $x_i$ from $x\in\mathcal{S}$ in proportion to $\beta(x)$.
\EndFor
\State Remove virtual transitions in $(\boldsymbol{t}',\boldsymbol{x}')$ and return $(\boldsymbol{t},\boldsymbol{x})$. 
\end{algorithmic}
\end{algorithm*}
Let $(\boldsymbol{t},\boldsymbol{x})$ denote a network path in $X$ fully compatible with $\boldsymbol{O}_k$, with $\boldsymbol{t}=\{t_1,\dots,t_n\}$ and $\boldsymbol{x}=\{x_1,\dots,x_n\}$; then, marginalizing over $(\boldsymbol{x}',\boldsymbol{u})$ the frame $\boldsymbol{t}'$ is independent of any observations and such that (cf. \cite{rao13a})
\begin{align*}
f_{\boldsymbol{t}'}(&t_1',\dots,t_m'|(\boldsymbol{t}, \boldsymbol{x}),\Omega,\boldsymbol{\theta},x_0)  =  \mathbb{I}(\boldsymbol{t} \subseteq \{ t_1',\dots,t_m'  \} ) \times  \\ 
& \prod_{i=0}^n(\Omega+Q_{x_i})^{\#(\{t_1',\dots,t_m'\}\cap (t_i,t_{i+1}))} e^{-(\Omega+Q_{x_i})(t_{i+1}-t_i)}
\end{align*}
with $t_0=0$ and $t_{n+1}=T$. Thus, it may be sampled in a collapsed step incorporating virtual transitions to times in $\boldsymbol{t}$, employing a succession of Poisson processes with rates $\{\Omega+Q_{x_i}$ : $x_i\in\boldsymbol{x}\}$. Note that $\boldsymbol{t}'$ along with $(\boldsymbol{t},\boldsymbol{x})$ induces preliminary sequences of missing observations in $\boldsymbol{O}_k'$ and uniformized transitions in $\boldsymbol{x}'$. Next, we target $\boldsymbol{u}|\boldsymbol{x}'$ sampling $m$ auxiliary-observation variables from \eqref{auxes}. 

Finally, we obtain a new path $(\boldsymbol{t},\boldsymbol{x})$ in full agreement with both real and auxiliary observations, producing $\boldsymbol{t},\boldsymbol{x},\boldsymbol{x}'$ in a blocked step. This simplifies to sampling a sequence $\boldsymbol{x}'|\boldsymbol{t}',\boldsymbol{u},\Omega,\boldsymbol{\theta},\boldsymbol{O}_k',x_0$ and removing virtual entries; it is achieved by employing dynamic arrays within a procedure for discrete-time state-space models as shown in Algorithm \ref{backForwAlgo}. Alternatively, note it is possible to employ a particle filtering approach within a forward procedure, in order to impose further memory constraints.

\subsubsection{Properties and considerations}
Along with observations and naturally restrictive constraints on state transitions within QNs, auxiliary variables in $\boldsymbol{u}$ allow us to limit the space a sampler is allowed to explore within each iteration. These restrictions apply both within forward and backward procedures and leave the underlying filtering equations unaltered, up to proportionality. Increasing the value of $p$ will make computationally expensive iterations less likely, at the cost of a higher dependence between subsequent realizations of $X$. Also, the term $(1-q)$ enters the forward procedure penalizing network paths with unobserved transitions and is only proportionally relevant when no observation exists.

\begin{figure*}[t]
\centering
\begin{tikzpicture}
\node at (0.5cm,4cm) {\small \textbf{Iteration} $\boldsymbol{n}$};
\draw[->,line width=0.20mm] (0cm,3cm) -- (15cm,3cm);
\draw (0,2.9cm) -- ++(0cm,0.2cm);

\node at (15.3cm,3cm) {$\boldsymbol{t}$};

\draw[-,dashed,color=black!80!white,line width=0.20mm] (1.9,4.58cm) -- ++(0cm,-7.7cm);
\draw[-,dashed,color=black!80!white,line width=0.20mm] (4.2,4.58cm) -- ++(0cm,-5.5cm);
\draw[-,dashed,color=black!80!white,line width=0.20mm] (10.3,4.58cm) -- ++(0cm,-5.5cm);
\draw[-,dashed,color=black!80!white,line width=0.20mm] (14.4,4.58cm) -- ++(0cm,-7.7cm);
\draw[-,dashed,color=black!80!white,line width=0.20mm] (1.9,5.08cm) -- ++(12cm,0cm);

\filldraw[black!60!white,draw=black] (1.9cm,3cm) circle [radius=0.09cm]; 
\filldraw[black!60!white,draw=black] (4.2cm,3cm) circle [radius=0.09cm]; 
\filldraw[black!60!white,draw=black] (5.7cm,3cm) circle [radius=0.09cm]; 
\filldraw[black!60!white,draw=black] (10.3cm,3cm) circle [radius=0.09cm]; 
\filldraw[black!60!white,draw=black] (12.1cm,3cm) circle [radius=0.09cm]; 
\filldraw[black!60!white,draw=black] (14.4cm,3cm) circle [radius=0.09cm]; 

\node[fill=white] at (1.93cm,2.6cm) {\footnotesize $1_{0\rightarrow 1}$};
\node[fill=white] at (4.23cm,2.6cm) {\footnotesize $2_{0\rightarrow 1}$};
\node[fill=white] at (5.73cm,2.6cm) {\footnotesize $1_{1\rightarrow 3}$};
\node[fill=white] at (10.33cm,2.6cm) {\footnotesize $1_{3\rightarrow 0}$};
\node[fill=white] at (12.13cm,2.6cm) {\footnotesize $2_{1\rightarrow 3}$};
\node[fill=white] at (14.43cm,2.6cm) {\footnotesize $2_{3\rightarrow 0}$};

\node[fill=white] at (7.2cm,5.08) {\small \textbf{Evidence}};
\filldraw[black!05!white,draw=black] (1.45cm,4.8cm) rectangle ++(0.9,0.6); 
\node at (1.9cm,5.08cm) {\small $\boldsymbol{1_{0\rightarrow \cdot}}$};
\filldraw[black!05!white,draw=black] (3.75cm,4.8cm) rectangle ++(0.9,0.6);
\node at (4.2cm,5.08cm) {\small $\boldsymbol{2_{0\rightarrow \cdot}}$};
\filldraw[black!05!white,draw=black] (9.85cm,4.8cm) rectangle ++(0.9,0.6); 
\node at (10.3cm,5.08cm) {\small $\boldsymbol{1_{\cdot\rightarrow 0}}$};
\filldraw[black!05!white,draw=black] (13.95cm,4.8cm) rectangle ++(0.9,0.6); 
\node at (14.4cm,5.08cm) {\small $\boldsymbol{2_{\cdot\rightarrow 0}}$};

\node[fill=white] at (1.07cm,1.5cm) {\small \textbf{Add virtual jumps}};
\draw[->,line width=0.20mm] (0cm,0.5cm) -- (15cm,0.5cm);
\draw (0,0.4cm) -- ++(0cm,0.2cm);

\node at (15.3cm,0.5cm) {$\boldsymbol{t}$};

\node[fill=white] at (1.93cm,0.1cm) {\footnotesize $1_{0\rightarrow 1}$};
\node[fill=white] at (4.23cm,0.1cm) {\footnotesize $2_{0\rightarrow 1}$};
\node[fill=white] at (5.73cm,0.1cm) {\footnotesize $1_{1\rightarrow 3}$};
\node[fill=white] at (10.33cm,0.1cm) {\footnotesize $1_{3\rightarrow 0}$};
\node[fill=white] at (12.13cm,0.1cm) {\footnotesize $2_{1\rightarrow 3}$};
\node[fill=white] at (14.43cm,0.1cm) {\footnotesize $2_{3\rightarrow 0}$};

\filldraw[black!60!white,draw=black] (1.9cm,0.5cm) circle [radius=0.09cm]; 
\filldraw[black!05!white,draw=black] (0.5cm,0.3cm) rectangle ++(0.4,0.4); 
\filldraw[black!00!white,draw=black] (0.7cm,0.5cm) circle [radius=0.09cm]; 
\filldraw[black!05!white,draw=black] (1.1cm,0.3cm) rectangle ++(0.4,0.4); 
\filldraw[black!00!white,draw=black] (1.3cm,0.5cm) circle [radius=0.09cm]; 
\filldraw[black!60!white,draw=black] (4.2cm,0.5cm) circle [radius=0.09cm]; 
\filldraw[black!05!white,draw=black] (3.4cm,0.3cm) rectangle ++(0.4,0.4); 
\filldraw[black!00!white,draw=black] (3.6cm,0.5cm) circle [radius=0.09cm]; 
\filldraw[black!05!white,draw=black] (5.5cm,0.3cm) rectangle ++(0.4,0.4); 
\filldraw[black!60!white,draw=black] (5.7cm,0.5cm) circle [radius=0.09cm]; 
\filldraw[black!05!white,draw=black] (4.5cm,0.3cm) rectangle ++(0.4,0.4); 
\filldraw[black!00!white,draw=black] (4.7cm,0.5cm) circle [radius=0.09cm]; 
\filldraw[black!60!white,draw=black] (10.3cm,0.5cm) circle [radius=0.09cm]; 
\filldraw[black!05!white,draw=black] (6.3cm,0.3cm) rectangle ++(0.4,0.4); 
\filldraw[black!00!white,draw=black] (6.5cm,0.5cm) circle [radius=0.09cm]; 
\filldraw[black!05!white,draw=black] (6.9cm,0.3cm) rectangle ++(0.4,0.4); 
\filldraw[black!00!white,draw=black] (7.1cm,0.5cm) circle [radius=0.09cm]; 
\filldraw[black!05!white,draw=black] (8.4cm,0.3cm) rectangle ++(0.4,0.4); 
\filldraw[black!00!white,draw=black] (8.6cm,0.5cm) circle [radius=0.09cm]; 
\filldraw[black!05!white,draw=black] (9.2cm,0.3cm) rectangle ++(0.4,0.4); 
\filldraw[black!00!white,draw=black] (9.4cm,0.5cm) circle [radius=0.09cm]; 
\filldraw[black!05!white,draw=black] (11.9cm,0.3cm) rectangle ++(0.4,0.4); 
\filldraw[black!60!white,draw=black] (12.1cm,0.5cm) circle [radius=0.09cm]; 
\filldraw[black!05!white,draw=black] (13.6cm,0.3cm) rectangle ++(0.4,0.4); 
\filldraw[black!00!white,draw=black] (13.8cm,0.5cm) circle [radius=0.09cm]; 
\filldraw[black!60!white,draw=black] (14.4cm,0.5cm) circle [radius=0.09cm]; 


\draw[-,dashed,color=black!80!white,line width=0.20mm] (0.7,0.3cm) -- ++(0cm,-1.3cm);
\draw[-,dashed,color=black!80!white,line width=0.20mm] (1.3,0.3cm) -- ++(0cm,-1.3cm);
\draw[-,dashed,color=black!80!white,line width=0.20mm] (3.6,0.3cm) -- ++(0cm,-3.4cm);
\draw[-,dashed,color=black!80!white,line width=0.20mm] (4.7,0.3cm) -- ++(0cm,-1.3cm);
\draw[-,dashed,color=black!80!white,line width=0.20mm] (5.7,0.3cm) -- ++(0cm,-3.4cm);
\draw[-,dashed,color=black!80!white,line width=0.20mm] (6.5,0.3cm) -- ++(0cm,-3.4cm);
\draw[-,dashed,color=black!80!white,line width=0.20mm] (7.1,0.3cm) -- ++(0cm,-3.4cm);
\draw[-,dashed,color=black!80!white,line width=0.20mm] (8.6,0.3cm) -- ++(0cm,-1.3cm);
\draw[-,dashed,color=black!80!white,line width=0.20mm] (9.4,0.3cm) -- ++(0cm,-3.4cm);
\draw[-,dashed,color=black!80!white,line width=0.20mm] (12.1,0.3cm) -- ++(0cm,-3.4cm);
\draw[-,dashed,color=black!80!white,line width=0.20mm] (13.8,0.3cm) -- ++(0cm,-3.4cm);

\node at (0cm,-1cm) {\large $\boldsymbol{u}$:};
\node[fill=white] at (0.7cm,-1cm) {\xmark};
\node[fill=white] at (1.3cm,-1cm) {\xmark};
\node[fill=white] at (1.9cm,-1cm) {\cmark};
\node[fill=white] at (3.6cm,-1cm) {\cmark};
\node[fill=white] at (4.2cm,-1cm) {\xmark};
\node[fill=white] at (4.7cm,-1cm) {\xmark};
\node[fill=white] at (5.7cm,-1cm) {\cmark};
\node[fill=white] at (6.5cm,-1cm) {\cmark};
\node[fill=white] at (7.1cm,-1cm) {\cmark};
\node[fill=white] at (8.6cm,-1cm) {\xmark};
\node[fill=white] at (9.4cm,-1cm) {\cmark};
\node[fill=white] at (10.3cm,-1cm) {\xmark};
\node[fill=white] at (12.1cm,-1cm) {\cmark};
\node[fill=white] at (13.8cm,-1cm) {\cmark};
\node[fill=white] at (14.4cm,-1cm) {\cmark};


\node[fill=white] at (2.01cm,-2.2cm) {\small \textbf{Empty frame, project evidence}};
\draw[->,line width=0.20mm] (0cm,-3.2cm) -- (15cm,-3.2cm);
\draw (0,-3.3cm) -- ++(0cm,0.2cm);

\node at (15.3cm,-3.2cm) {$\boldsymbol{t}$};

\filldraw[black!00!white,draw=black] (1.9cm,-3.2cm) circle [radius=0.09cm]; 
\filldraw[black!00!white,draw=black] (3.6cm,-3.2cm) circle [radius=0.09cm]; 
\filldraw[black!60!white,draw=black] (4.2cm,-3.2cm) circle [radius=0.09cm]; 
\filldraw[black!00!white,draw=black] (5.7cm,-3.2cm) circle [radius=0.09cm]; 
\filldraw[black!00!white,draw=black] (6.5cm,-3.2cm) circle [radius=0.09cm]; 
\filldraw[black!00!white,draw=black] (7.1cm,-3.2cm) circle [radius=0.09cm]; 
\filldraw[black!00!white,draw=black] (9.4cm,-3.2cm) circle [radius=0.09cm]; 
\filldraw[black!60!white,draw=black] (10.3cm,-3.2cm) circle [radius=0.09cm]; 
\filldraw[black!00!white,draw=black] (12.1cm,-3.2cm) circle [radius=0.09cm]; 
\filldraw[black!00!white,draw=black] (13.8cm,-3.2cm) circle [radius=0.09cm]; 
\filldraw[black!00!white,draw=black] (14.4cm,-3.2cm) circle [radius=0.09cm]; 

\draw[-,dashed,color=black!80!white,line width=0.20mm] (4.2,-3.3cm) -- ++(0cm,-2.4cm);
\draw[-,dashed,color=black!80!white,line width=0.20mm] (10.3,-3.3cm) -- ++(0cm,-2.4cm);

\node[fill=white] at (4.23cm,-3.6cm) {\footnotesize $2_{0\rightarrow 1}$};
\node[fill=white] at (10.33cm,-3.6cm) {\footnotesize $1_{3\rightarrow 0}$};

\node[fill=white] at (1.32cm,-4.7cm) {\small \textbf{FFBS: iteration $\boldsymbol{n+1}$}};
\draw[->,line width=0.20mm] (0cm,-5.7cm) -- (15cm,-5.7cm);
\draw (0,-5.8cm) -- ++(0cm,0.2cm);

\node at (15.3cm,-5.7cm) {$\boldsymbol{t}$};

\filldraw[black!60!white,draw=black] (1.9cm,-5.7cm) circle [radius=0.09cm]; 
\filldraw[black!60!white,draw=black] (4.2cm,-5.7cm) circle [radius=0.09cm]; 
\filldraw[black!60!white,draw=black] (6.5cm,-5.7cm) circle [radius=0.09cm]; 
\filldraw[black!60!white,draw=black] (7.1cm,-5.7cm) circle [radius=0.09cm]; 
\filldraw[black!60!white,draw=black] (10.3cm,-5.7cm) circle [radius=0.09cm]; 
\filldraw[black!60!white,draw=black] (14.4cm,-5.7cm) circle [radius=0.09cm]; 

\node[fill=white] at (1.93cm,-6.1cm) {\footnotesize $1_{0\rightarrow 2}$};
\node[fill=white] at (4.23cm,-6.1cm) {\footnotesize $2_{0\rightarrow 1}$};
\node[fill=white] at (6.44cm,-6.1cm) {\footnotesize $1_{2\rightarrow 3}$};
\node[fill=white] at (7.22cm,-6.1cm) {\footnotesize $2_{1\rightarrow 3}$};
\node[fill=white] at (10.33cm,-6.1cm) {\footnotesize $1_{3\rightarrow 0}$};
\node[fill=white] at (14.43cm,-6.1cm) {\footnotesize $2_{3\rightarrow 0}$};

\end{tikzpicture}
\vspace{10pt}
\caption{Task transition diagram with a single iteration in the proposed sampler, for a bottleneck network with three servers. Here, tasks 1 and 2 are observed entering and leaving the network. First, start with a path whose task transitions are fully compatible with the evidence. Then, supplement it with virtual jumps at the corresponding rates, and produce auxiliary variables across real and virtual nodes. Next, empty the uniformized frame and propagate both real and auxiliary evidence; imposing task transitions or virtual jumps within clamped nodes. Finally, repopulate the frame via forward filtering backward sampling, hence maintaining agreement with the existing evidence.} \label{samplerIter}
\end{figure*}
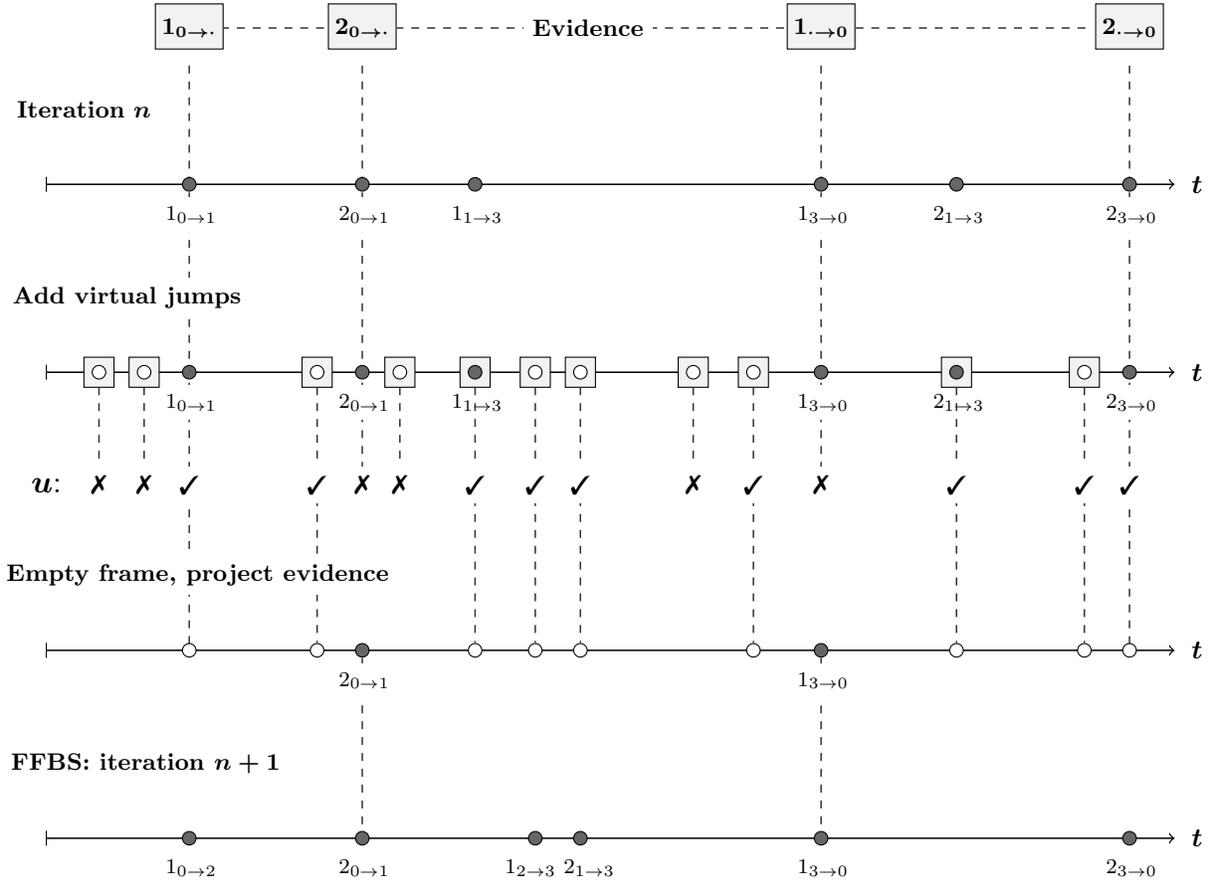
In Figure \ref{samplerIter} we observe a task transition diagram with a single iteration in the proposed sampler, for the bottleneck network in Figure \ref{exmplQueues} (top). In this example, two tasks (numbered 1 and 2) are observed entering and leaving the network at different times; however, there exists no information regarding job services within the network. In each iteration, the sampler begins with a network path whose task transitions are fully compatible with the existing evidence. In an initial step, the existing path is supplemented with virtual transitions at the corresponding Poisson rates. In the Figure, we observe that nodes for both virtual jumps and the unobserved job services are superimposed over shaded boxes; the boxes represent further evidence for the lack of task arrivals or departures at these times. Next, auxiliary variables are produced across real and virtual jumps, the subsets are loosely represented by ticks ($\Gamma\cup\varnothing$) and crosses ($\{\mathcal{T}(x'_{i-1},x'_i)\}$) for open and clamped nodes respectively. Then, the uniformized frame is emptied and both real and auxiliary evidence is propagated, imposing task transitions or virtual jumps within clamped nodes and resulting in a restricted frame for possible network paths. Finally, a new compatible path is sampled via forward filtering backward sampling as summarized in Algorithm \ref{backForwAlgo}; this will consider the imposed task transitions and weight successive network states over the clamped epochs. The resulting path is fully compatible with the observed evidence, however, notice that task transitions at arrival or departure times may change between iterations.

Note that by choosing $\Omega$ strictly greater than all values in the diagonal of $Q$, the resulting Markov chain over posterior network transitions is irreducible. Increasing the dominating rate will improve mixing in exchange for higher computational requirements. Finally, we note that a high value of $p$ may hinder the sampler from fully exploring the posterior range of network paths.

\subsubsection{Parameter sampling}
Finally, given a new family of network realizations $\boldsymbol{X}=\{X^k\}_{k=1,\dots,K}$ fully compatible with observation sequences $\tilde{\boldsymbol{O}}=\{\boldsymbol{O}_k\}_{k=1,\dots,K}$, we may obtain posterior samples of arrival and service rate parameters. For traditional FCFS stations this is such that
$$\lambda_c | \boldsymbol{X} \sim \text{Gamma}\big( \textstyle\delta_c, \sum_{k=1}^K T_k \big)$$ and  
$$\mu^c_i | \boldsymbol{X} \sim \text{Gamma}\big( \gamma^c_i, \tau^c_i \big),$$
for $c\in\mathcal{C}, i=1,\dots,M$; assuming independent network parameters and uninformative priors. Here $\delta_c, \gamma^c_i$ and $\tau^c_i$ denote respectively the number of class $c$ arrivals, class $c$ jobs served at station $i$ and the time server $i$ has been occupied by a class $c$ job, in all realizations in $\boldsymbol{X}$. Finally, posterior probability vectors for class $c$ routings in every node $i=1,\dots,M$ are given by
$$P^c_{i,\cdot} | \boldsymbol{X} \sim \text{Dir}\big( \boldsymbol{1} + \boldsymbol{\kappa}^c_i  \big)$$
where $\boldsymbol{\kappa}^c_i$ defines a vector of transition counts from server $i$ in $\boldsymbol{X}$. Arrival posteriors in $\boldsymbol{p}$ are defined the same way. We note that in order to ease identifiability in the inferential problem, it is also possible to fix parameters, incorporate conjugate priors or to impose inequality constraints and bounds across parameters; we will show examples in Section \ref{Examps} below. Also, the above expressions must be altered when stations respond to prioritization regimes other than FCFS (see Example 3 in Section \ref{Examps}).

\section{Examples} \label{Examps}

In the following, we discuss results obtained across three example networks with both synthetic and real data, in order of increasing difficulty. In all cases, results are obtained through a JAVA implementation of the proposed sampler, and starting compatible network paths have been manually assigned.

The examples demonstrate the ability of the proposed algorithm in order to handle missing data in multi-class inferential problems with varying service disciplines, class switching and imposed prior constraints. Hence, the sampler offers the means to overcome necessary assumptions linked to the common use of product form equilibrium expressions for QNs. To the best of our knowledge, there exists no alternative approach overcoming these restrictions when drawing exact inference in general open Markovian networks. 

\begin{figure*}[t]
  \centering
    \includegraphics[width=\textwidth]{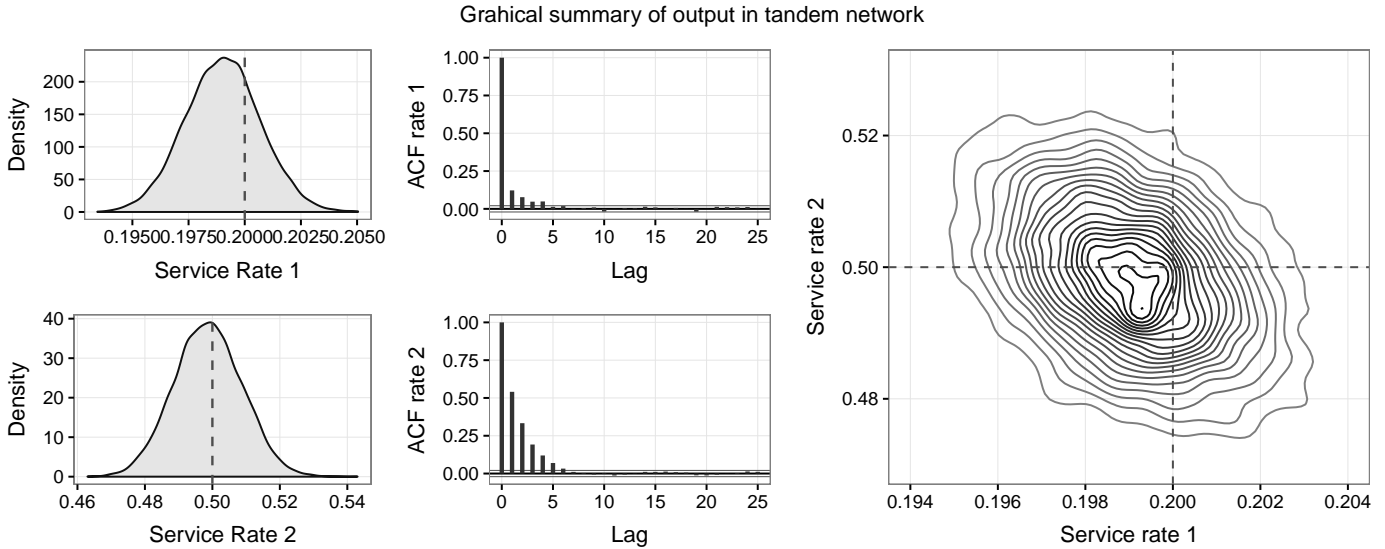}   
   \caption{Graphical summary of output for a single-class tandem network fitted to synthetic data. On the left, we observe the posterior distribution of service rate parameters with original values $\mu_1=0.2$ and $\mu_2=0.5$. Black bars in the center correspond to autocorrelation values. On the right, a contour plot for the joint posterior density of rates (dashed lines represent real values).} \label{CorTandem}
\end{figure*}
\subsection{Tandem network} In the simplest example, we analyse simulated data on a tandem network with two $M/M/1$ stations, FCFS service disciplines and a single task class. Data is generated so that true service rates are $\mu_1=0.2$ and $\mu_2=0.5$, arrivals are given by $\lambda=0.12$ and the network topology is defined by a routing probability matrix $P$ such that $P_{1,2} = 1$ and $P_{2,0}=1$. Also, jobs enter directly into the first queue and $p_{0,1} = 1$. 

For the inferential problem, job service observations (in first station) are always ignored and the only source of information are end-to-end measurements. Thus, available knowledge is limited to the times when tasks enter the queue on the first station and when they depart through the second station. Overall, we examine 5000 realizations totalling 17827 tasks during 115601 time units. For the purpose, the network topology in $P$ is fixed deterministic, since there exists a unique route from start to completion of tasks. Also, in order to ensure identifiability we impose an inequality constraint on service rates and assign fairly uninformative parameter priors, so that $$\pi(\mu_1,\mu_2)\propto \mathbb{I}(\mu_1\leq\mu_2)\times\exp(-10^{-3}(\mu_1+\mu_2)).$$
Note that the problem directly links to the inferential task with two exponentially distributed random variables when only its sum is observed, with the further complexity that unknown waiting times have to be discounted from the empirical observations.

In Figure \ref{CorTandem} (right) we observe a contour plot for the joint posterior kernel density estimation over service rates, and we notice a significant negative correlation in values (the dashed vertical and horizontal lines represent the original parameters values in the network path simulations).  Results are obtained across two chains with 100000 iterations each, a 10000 burn-in stage, varying starting rates and different scales for dominating rates and probabilities producing auxiliary-observations, so that $p_1=0$ and $\Omega_1=2\max_{x\in\mathcal{S}}|Q_x|$ and $p_2=0.25, \Omega_2=1.5\max_{x\in\mathcal{S}}|Q_x|$. Note that the second chain is produced employing restrictive auxiliary-observations as opposed to the first; hence, stronger serial dependencies across subsequent latent paths in the network should be expected. Yet, the remainder plots show marginal posterior kernel density estimations for both service rates, along with an autocorrelation summary across a thinned sample in the second chain, showing a satisfactory mixing. 

A discussion on the effects and computational gains resulting from employing restrictive auxiliary observations follows in the next example. In general, networks of interest are complex and $p=0$ would pose a computationally infeasible problem. Also, even in simple networks such as this example, computing times can be excessive, and considerable reductions can be traded at the cost of higher serial dependences.

\begin{table*}[t]
\caption{Summary statistics for posterior service rates along with computing times across three chains tuned differently in the bottleneck network in Figure \ref{exmplQueues} (top).} \label{summaryMCMC}
\renewcommand{\arraystretch}{1.3}
\setlength{\tabcolsep}{17pt}
\centering
\begin{tabular}{lllllllll}
\hline
      & Real & \multicolumn{2}{l}{Summary} & \multicolumn{5}{l}{Quantiles}      \\ \cline{3-9} 
    &  & Mean        & StDev      & 2.5\%   & 25\%    & 52\%    & 75\%    & 97.5\%     \\ \hline
$\mu_1^1$ & \textbf{0.3}  & 0.311       & 0.019         & 0.273 & 0.297 & 0.310 & 0.323 & 0.350   \\
$\mu_1^2$ & \textbf{0.25} & 0.237       & 0.017         & 0.204 & 0.225 & 0.236 & 0.247 & 0.271       \\
$\mu_1^3$ & \textbf{0.2}  & 0.185       & 0.016         & 0.154 & 0.173 & 0.184 & 0.196 & 0.218           \\
$\mu_2^1$ & \textbf{0.7}  & 0.709       & 0.052         & 0.613 & 0.673 & 0.707 & 0.743 & 0.818     \\
$\mu_2^2$ & \textbf{0.5}  & 0.537       & 0.040         & 0.462 & 0.509 & 0.536 & 0.564 & 0.620      \\
$\mu_2^3$ & \textbf{0.3}  & 0.297       & 0.028         & 0.245 & 0.277 & 0.296 & 0.315 & 0.354     \\
$\mu_3^1$ & \textbf{1.5}  & 1.626       & 0.088         & 1.458 & 1.566 & 1.625 & 1.684 & 1.802    \\
$\mu_3^2$ & \textbf{1.2}  & 1.206       & 0.069         & 1.074 & 1.159 & 1.204 & 1.252 & 1.346     \\
$\mu_3^3$ & \textbf{0.8}  & 0.728       & 0.052         & 0.630 & 0.692 & 0.727 & 0.762 & 0.833    \\ \hline \addlinespace[0.1cm] \hline
 & $p$       & \multicolumn{2}{l}{$\Omega$}  & \multicolumn{2}{l}{Run Time} &  \textbf{ESS:} & Mean & Min    \\ \cline{1-9} 
1 & 0.7       & \multicolumn{2}{l}{$2\max_{x\in\mathcal{S}}|Q_x|$}  & \multicolumn{2}{l}{6056.5s}  & & 14575 & 7246   \\
2 & 0.5       & \multicolumn{2}{l}{$1.5\max_{x\in\mathcal{S}}|Q_x|$}  & \multicolumn{2}{l}{24709.4s} & & 26069 & 14750   \\
3 & 0.2       & \multicolumn{2}{l}{$1.2\max_{x\in\mathcal{S}}|Q_x|$}  & \multicolumn{2}{l}{73993.2s} & & 42101 & 22460  \\ \hline
\end{tabular}
\end{table*}
\subsection{Bottleneck network} We examine simulated data in the bottleneck network in Figure \ref{exmplQueues} (top), with 3 FCFS stations and 3 different task classes. The true service rates can be observed in Table \ref{summaryMCMC}, and task arrivals are given by  $\lambda_1=0.08, \lambda_2=0.06$ and $\lambda_3=0.04$. In this case, along with end-to-end measurements, approximately half of all generated job service observations are retrieved so that $q=0.5$. The network topology is defined by $\{P^c,p^c:c\in\mathcal{C}\}$, where
$$P^c= \bordermatrix{  & \  0 \ &   1 \ &   2 \ &   3 \ \cr
                             1 & \ 0 \ & 0 \ & 0 \ & 1 \ \cr
                             2 & \ 0 \ & 0 \ & 0 \ & 1 \ \cr
                             3 & \ 1 \ & 0 \ & 0 \ & 0 \ },$$
is identical for all three classes and assumed to be known. In addition, job entries are split evenly, i.e. $p^c_{0,1} = 0.5$ and $p^c_{0,2} = 0.5$ for all $c\in\mathcal{C}$. 

In total, we analyse 500 network realizations totalling 1281 tasks during 5083 time units. In order to ease identifiability we assume the existence of a slow, medium and fast server; and assign rather uninformative parameter priors, i.e.
\begin{align*}
\pi(\mu^c_1,&\mu^c_2,\mu^c_3) \\ 
& \propto \mathbb{I}(\mu^c_1\leq\mu^c_2\leq\mu^c_3)\times\exp(-10^{-3}(\mu^c_1+\mu^c_2+\mu^c_3))
\end{align*}
for all $c\in\mathcal{C}$. Note that this network type may not be analysed by means of product-form representations centred around figures of queue-lengths (c.f \cite{Wang2016}). This is because traditional BCMP networks require FCFS stations to share service rates across task classes. On the other hand, an MCMC sampler as presented in \citep{sutton2011} can be extended in order to handle general service distributions and target network path transitions; however, the framework is not designed for such aim, it would require an additional Metropolis-Hastings step and it is likely to perform poorly.

\begin{table}[h!]
\caption{Correlation matrix between service rate parameters in a bottleneck network.}
\label{corMat}
\renewcommand{\arraystretch}{1.3}
\setlength{\tabcolsep}{4pt}
\centering
\begin{tabular}{crrrrrrrr}
\hline
     & $\mu_1^2$  & $\mu_1^3$  & $\mu_2^1$  & $\mu_2^2$  & $\mu_2^3$  & $\mu_3^1$  & $\mu_3^2$  & $\mu_3^3$  \\
\hline
$\mu_1^1$ & -0.02 & -0.01 & -0.05 & 0.01  & 0.01  & -0.05 & 0.01  & 0.02  \\
$\mu_1^2$ &       & -0.02 & 0.00  & -0.03 & 0.01  & 0.01  & -0.04 & 0.01  \\
$\mu_1^3$ &       &       & 0.01  & 0.01  & -0.09 & 0.00  & -0.01 & -0.06 \\
$\mu_2^1$ &       &       &       & -0.03 & -0.05 & -0.13 & 0.01  & 0.01  \\
$\mu_2^2$ &       &       &       &       & -0.02 & 0.00  & -0.10 & 0.01  \\
$\mu_2^3$ &       &       &       &       &       & 0.00  & 0.00  & -0.10 \\
$\mu_3^1$ &       &       &       &       &       &       & 0.01  & 0.01  \\
$\mu_3^2$ &       &       &       &       &       &       &       & 0.01 \\ \hline
\end{tabular}
\end{table}
In Table \ref{summaryMCMC} we observe summary statistics, computing times and effective sample sizes across three chains with 100000 iterations each, a 10000 burn-in stage, varying starting rates and different scales for dominating rates $\Omega$ and probabilities $p$ producing auxiliary-observations. There, we notice a good trade-off between effective samples and the drastic decrease in computing times required when imposing strong serial relations on network paths across subsequent iterations in the sampler. This is the case even when the volume of virtual jumps produced is reduced, and emphasizes the need for such slice sampling techniques in inferential problems with QNs. In addition, Table \ref{corMat} displays the overall posterior correlation matrix between service rate parameters, and shows very mild relations in rates for each task class. There, we notice the importance of employing posterior samples from the produced chains in order to answer extrapolation-type questions in network systems. Finally, note it is possible to ease imposed restrictions on the network topology and to employ different service disciplines across servers (see next example).

\subsection{Feedback network} Finally, we show how the proposed sampling scheme may be used to analyse a real data set. For the purpose, we employ work-logs for medical clinicians. Briefly, the data set includes task requests and completions for individual doctors outside the 9:00-17:00 Monday to Friday \textit{in hours} settings. It belongs to two jointly coordinated university hospitals in the United Kingdom, together servicing a geographical region with over 2.5 million residents. 
\begin{figure}[h]
\centering
\begin{tikzpicture}
\draw[->,line width=0.20mm] (0cm,0cm) -- (7.6cm,0cm);
\draw (0,-0.1cm) -- ++(0cm,0.2cm);
\draw (1.2,-0.1cm) -- ++(0cm,0.2cm);
\draw (2.4,-0.1cm) -- ++(0cm,0.2cm);
\draw (2.9,-0.1cm) -- ++(0cm,0.2cm);
\draw (4.2,-0.1cm) -- ++(0cm,0.2cm);
\draw (4.8,-0.1cm) -- ++(0cm,0.2cm);
\draw (6.9,-0.1cm) -- ++(0cm,0.2cm);
\node at (7.9cm,0cm) {$\boldsymbol{t}$};
\node at (0.00cm,-0.3cm) {\footnotesize $0$};
\node at (1.21cm,-0.3cm) {\footnotesize $t_1$};
\node at (2.41cm,-0.3cm) {\footnotesize $t_2$};
\node at (2.91cm,-0.3cm) {\footnotesize $t_3$};
\node at (4.21cm,-0.3cm) {\footnotesize $t_4$};
\node at (4.81cm,-0.3cm) {\footnotesize $t_5$};
\node at (6.91cm,-0.3cm) {\footnotesize $t_6$};
\draw[-,dashed,line width=0.20mm] (1.2cm,0cm) -- ++ (0cm,0.75cm);
\draw[-,dashed,line width=0.20mm] (2.4cm,0cm) -- ++ (0cm,1.5cm);
\draw[-,dashed,line width=0.20mm] (2.9cm,0cm) -- ++ (0cm,2.25cm);
\draw[-,dashed,line width=0.20mm] (4.2cm,0cm) -- ++ (0cm,0.75cm);
\draw[-,dashed,line width=0.20mm] (4.8cm,0cm) -- ++ (0cm,2.25cm);
\draw[-,dashed,line width=0.20mm] (6.9cm,0cm) -- ++ (0cm,1.5cm);
\draw[black!70!white,line width=1.3mm] (1.2cm,0.75cm) -- (4.2cm,0.75cm);
\node at (0.75cm,0.75cm) {\footnotesize \textbf{Fall}};
\draw[black!70!white,line width=1.3mm] (2.4cm,1.5cm) -- (6.9cm,1.5cm);
\node at (1.6cm,1.5cm) {\footnotesize \textbf{Clerking}};
\draw[black!70!white,line width=1.3mm] (2.9cm,2.25cm) -- (4.8cm,2.25cm);
\node at (2.15cm,2.25cm) {\footnotesize \textbf{Urgency}};
\end{tikzpicture}
\caption{Sample diagram with a subset of tasking data linked to a clinician during a shift} \label{sampleData}
\end{figure}
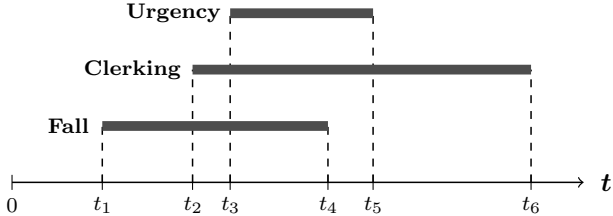
In Figure \ref{sampleData} we show a diagram with a small of subset of data linked to a clinician during a shift; there we observe three overlapping tasks recorded in the system (from request to completion), and each belonging to a different class. Note that it is not possible to know when the clinician was engaged with each duty; as individual jobs for tasks are not registered when queueing or being routed across teams of administrative staff, nurses and doctors. An extended description of the data set may be found in \cite{Perez201634}.

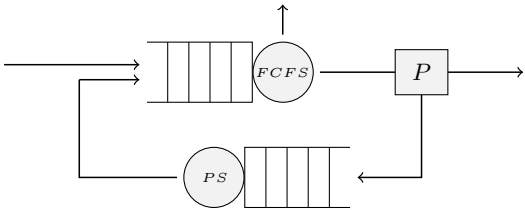
\begin{figure}[h]
\centering
\begin{tikzpicture}
\draw (7,0) -- ++(1.4cm,0) -- ++(0,-0.8cm) -- ++(-1.4cm,0);
\foreach \i in {1,...,4}
  \draw (8.4cm-\i*8pt,0) -- +(0,-0.8cm);
\filldraw[fill=black!05!white] (8.4cm+0.41cm,-0.4cm) circle [radius=0.4cm]; 

\draw (9.7,-1.4cm) -- ++(-1.4cm,-0cm) -- ++(0cm,-0.8cm) -- ++(1.4cm,0);
\foreach \i in {1,...,4}
  \draw (8.3cm+\i*8pt,-1.4cm) -- +(0,-0.8cm);
\filldraw[fill=black!05!white] (8.3cm-0.41cm,-1.4cm-0.4cm) circle [radius=0.4cm]; 

\filldraw[fill=black!05!white] (10.3cm,-0.1cm) rectangle (11cm,-0.7cm);

\draw[-,line width=0.2mm] (9.3cm,-0.4cm) -- ++(1,0.0cm);
\draw[->,line width=0.20mm] (8.805,0.1cm) -- ++(0.0,0.4cm);
\draw[->,line width=0.20mm] (11,-0.4cm) -- ++(1,0cm);
\draw[->,line width=0.20mm] (10.65cm,-0.7cm) -- ++(0cm,-1.1cm) -- ++(-0.85cm,0cm);
\draw[->,line width=0.20mm] (6.1cm,-0.5cm) -- ++(0.8,0.0cm);
\draw[<-,line width=0.20mm] (6.9cm,-0.3cm) -- ++(-1.8cm,0cm);
\draw[line width=0.20mm] (7.4cm,-1.8cm) -- ++(-1.3cm,0cm) -- ++(0cm,1.3cm);

\node[align=center] at (10.65cm,-0.4cm) {\small $P$};
\node[align=center] at (8.8cm,-0.4cm) {\tiny $FCFS$};
\node[align=center] at (7.9cm,-1.8cm) {\tiny $PS$};
\end{tikzpicture}
\vspace{10pt}
\caption{Feedback network with two M/M/1 servers and route switching among task classes.} \label{feedbackQ}
\end{figure}

Multiple tasks are grouped across 14 categories and analysed with a \textit{feedback network} as shown in Figure \ref{feedbackQ}. There, we notice the presence of two M/M/1 servers with alternative disciplines and route switching among classes. Task observations for doctors are of roughly two kinds, based on whether they require \textit{engagement} or not. Many tasks are recorded and erased within doctor work-logs in a short time span, due to no need for action; on the other hand, the remainder of tasks exhibit long processing times indicating the need for considerable doctor activity.

In the proposed example, arrival jobs are buffered within an administrative FCFS priority type queue and depart to a transition center where they either leave the system or get routed for processing with some unknown probability. Once they are assigned to further processing, they join the doctor's processing centre and switch their routing mechanism; so they will depart the network next time they undergo administrative processing in the first queue. The service station aimed to capture strain on doctor workload is assigned a \textit{processor sharing} (PS) discipline with a single worker, aiming to accommodate doctors attending concurrent duties outside standard working hours. No job service observations are available, so that $q=0$ and only the arrival and departure times for tasks to the network are observed.

\begin{table}[b!]
\caption{Summary statistics for rate and routing parameters in a feedback network.} \label{summaryMCMC2pars}
\renewcommand{\arraystretch}{1.3}
\setlength{\tabcolsep}{12pt}
\centering
\begin{tabular}{lllll}
\hline
     &   & \multicolumn{2}{l}{Summary}   \\ \cline{3-4} 
     &  & Mean        & StDev        \\ \hline 
Buffer & $\mu_1$       & 3.397         & 0.072   \\
Admission & $\mu_2^1$        & 0.587         & 0.038    \\
 & $P^1_{1,0}$        & 0.250          & 0.026       \\
Certification death & $\mu_2^2$        & 0.723         & 0.159   \\
 & $P^2_{1,0}$        & 0.601          & 0.061  \\
Check patient & $\mu_2^{12}$        & 0.650          & 0.051      \\
 & $P^{12}_{1,0}$        & 0.399          & 0.033   \\
Clerking & $\mu_2^3$        & 0.731         & 0.018   \\
 & $P^3_{1,0}$        & 0.061         & 0.007    \\
Clinical review & $\mu_2^4$        & 0.608          & 0.015     \\
 & $P^4_{1,0}$        & 0.333          & 0.014  \\
Address relatives & $\mu_2^5$        & 0.361          & 0.079    \\
 & $P^5_{1,0}$        & 0.470          & 0.098     \\
Drug prescribing & $\mu_2^6$        & 0.920          &  0.028   \\
 & $P^6_{1,0}$        & 0.547          & 0.012     \\
Early warning & $\mu_2^7$        & 0.488          & 0.017   \\
 & $P^7_{1,0}$        & 0.346          & 0.017     \\
Fall of patient & $\mu_2^8$        & 0.575          & 0.080     \\
 & $P^8_{1,0}$        & 0.318          & 0.054   \\
None of above & $\mu_2^{10}$        & 0.339         & 0.013     \\
 & $P^{10}_{1,0}$        & 0.393          & 0.018     \\
Other services & $\mu_2^9$        & 0.209          & 0.012   \\
 & $P^9_{1,0}$        & 0.435          & 0.024    \\
Procedure request & $\mu_2^{11}$        & 0.635          & 0.027 \\
 & $P^{11}_{1,0}$        & 0.387          & 0.019     \\
Test request & $\mu_2^{13}$        & 0.565          & 0.013   \\
 & $P^{13}_{1,0}$        & 0.263          & 0.011    \\
Urgent response & $\mu_2^{14}$        & 0.209          & 0.028      \\
 & $P^{14}_{1,0}$        & 0.306          & 0.059    \\ \hline 
\end{tabular}
\end{table}
In total, we analyse a reduced subset of 10000 doctor shifts roughly distributed across 4 years of observations. The network topology is partially known; i.e. $P^c_{2,1} = p^c_{0,1} = 1$ for all task classes, and $P^c_{1,0} = 1$ after tasks have undergone processing and hence switched routing mechanism. However, $P^c_{1,0} = 1 - P^c_{1,2}$ needs to be determined for all existing task classes. Processing rates for tasks are assumed equal in the first service station and different in the PS server; we assign no constraints and we impose loosely uninformative priors such that
$$\pi(\mu^c_i) \propto \exp(-10^{-3}\mu^c_i)$$
for all $i\in\{1,2\}$ and $c\in\mathcal{C}$. Also, note that within a PS discipline posterior rates given network realizations are given by
$$\mu^c | \boldsymbol{X} \sim \text{Gamma}\big( \textstyle \gamma^c, \sum_{k=1}^K \int_{0}^{T_k}  \phi_c^k(t) \mathrm{d}t  \big),$$
for all $c\in\mathcal{C}$, where $\gamma^c$ denotes the number of class $c$ jobs served at the station in all realizations in $\boldsymbol{X}$; and
$$\phi_c^k(t) = \frac{\textstyle \sum_{j=0}^{J} \mathbb{I}(\text{Job $j$ is class $c$}) \cdot \mathbb{I}(a_j < t < d_j)}{\textstyle \sum_{j=0}^{J} \mathbb{I}(a_j < t < d_j)},$$
where summations are across all jobs processed in the PS station in realization $k$, and $a_j,d_j$ denote the arrival and departure times of the job to the server. In Table \ref{summaryMCMC2pars} we observe summary statistics for parameters across two chains with 100000 iterations each, a 50000 burn-in stage and varying starting rates. In one chain, we use $\Omega=2\max_{x\in\mathcal{S}}|Q_x|$ and $p = 0.75$; in the second we have $\Omega=1.5\max_{x\in\mathcal{S}}|Q_x|$ and $p = 0.5$.

\begin{table}[h!]
\caption{Point estimates and standard errors for average processing times excluding waiting times, across different tasks. These relate to times from entry to departure in network.} \label{summaryMCMC2}
\renewcommand{\arraystretch}{1.3}
\setlength{\tabcolsep}{5pt}
\centering
\begin{tabular}{llllll}
\hline
      & \multicolumn{2}{l}{Completion} &       & \multicolumn{2}{l}{Completion} \\ \cline{2-3} \cline{5-6} 
         & Mean    & StErr &          & Mean    & StErr \\ \hline
Admission & 1.796 & 0.087 & Early w. & 1.828  & 0.046   \\
Cert. death & 0.985 & 0.141 & Fall & 1.701  & 0.176  \\
Check  & 1.400  & 0.073  & None   & 2.263  & 0.076  \\
Clerking & 1.856  & 0.032  & Other &  3.173  & 0.173   \\
Cl. review & 1.588  &0.027 &  Procedure &  1.442  & 0.045  \\
Address rel. & 1.971 & 0.374 & Test & 1.818  & 0.032  \\
Drug pres. & 0.920  & 0.017 & Urgent & 3.877  & 0.491  \\ \hline 
\end{tabular}
\end{table}
In addition, Table \ref{summaryMCMC2} shows point estimates and standard errors for average completion times in all task types, these correspond to the full processing times from entry to departure in the network (excluding queueing times) and are reported in hour units. Hence, we notice it is possible to assess workload both globally and across single components in the system, thus allowing to answer extrapolation kinds of questions on workload; i.e. in relation to means, variances and extreme values for system strain under likely alterations.

\section{Discussion} \label{Discussion}
This paper has presented a flexible approach for carrying exact Bayesian inference within known or hypothesized queueing networks. Its focus is on multi-class, open and Markovian cases and the approach is centred around the underlying continuous-time Markov chains induced by these complex stochastic systems. The proposed method relies on a slice sampling technique with mappings to the space of task transitions across servers in the network. It sits well over uniformization-oriented MCMC approaches introduced in \cite{rao13a} and can deal with missing data, imposed prior knowledge and strong serial dependencies posing a complex inferential task (cf. \cite{sutton2011}).

The need for such inferential frameworks with missing data is justified by the ability of general-form networks to allow evaluating response times in complex systems. Overall, recovering measures such as processing times is a technically difficult task when designing increasingly complex IT systems \citep{Liu200636}, or in service delivery networks (such as those in hospitals) due to ethical issues with such an intrusive process \citep{Perez201634}. Yet, QNs provide the tools to assess system alterations, diagnose poor performance or evaluate robustness to spikes in workload.

The advantage of the presented inferential method is that it permits retrospectively assessing the likely status of systems at any point in time; rather than only providing summary information on strain over individual bits. However, limitations relate to tractability restrictions with high-magnitude networks. In such cases, controlling the dimensionality of unobservable state spaces requires imposing strong serial dependencies within simulated latent network paths across steps in the sampler. This however may restrict the produced chain from exploring the posterior range of network paths efficiently. Approximate inferential frameworks relying on reduced product-form simplifications of state beliefs may improve the scalability of the method. Moreover, it is possible to explore the use of particle filtering approaches along with auxiliary variables for this purpose, since clamping explorable spaces within filtering procedures would likely ease the usual challenges regarding particle degeneracy; that is, ending with a very few particles having non-zero weights. 

Also, the use of the uniformization technique will limit applications of the present framework to the study of purely Markovian processes. While it is possible to employ Markov-modulated regimes that adapt service and arrival rates to network states, this will greatly expand state spaces under consideration. Also, uniformization may deem the sampler computationally inefficient should service rates vary greatly across queues or job classes, as certain transition types will greatly dominate the underlying discrete time Markov chain.

Finally, the paper assumes that the volume of job service observations retrieved across the network is given by $\%(100\cdot q)$ of the total processing during a fixed time interval. For simplicity, $q$ is assumed fixed and known to the user. Many network structures (such as bottleneck networks) will allow for uncertainty regarding this parameter to be quantified by means of the presented sampler, as each iteration will provide a total number of network transitions complementing the observation number as a sufficient statistic. However, it is necessary to impose the knowledge of $q$ in order to ensure model identifiability whenever networks contain either global or self-loops.

\section*{Supplementary material}

Synthetic data along with a Java implementation of the algorithm can be found in \url{https://bitbucket.org/ikertxo1986/auxvarsamplerjava} or \url{https://github.com/IkerPerez/auxVarSampler}. This allows to reproduce the results within the examples above.

\section*{Acknowledgements}

We would like to thank the anonymous reviewers for their valuable remarks and suggestions that have improved the quality of this paper.

\bibliographystyle{apa}     
\bibliography{bibliography}

\end{document}